\documentclass[reprint,aps,pra,superscriptaddress,showpacs,amsmath
,amssymb]{revtex4-1}
\usepackage{mathtools}
\usepackage{graphicx}
\usepackage{dcolumn}
\usepackage{bm}
\usepackage{multirow}

\begin{document}
\newcommand{\pd}[2]{\frac{\partial #1}{\partial #2}} 
\newcommand{\ket}[1]{\left| #1 \right>} 
\newcommand{\bra}[1]{\left< #1 \right|} 
\newcommand{\braket}[2]{\left< #1 \vphantom{#2} \right|
 \left. #2 \vphantom{#1} \right>} 
\renewcommand{\b}{\mathbf}
\newcommand{\sech}{\text{sech}}

\title{Vector-Soliton Storage and Three-Pulse Area Theorem}

\author{Rodrigo Guti\'{e}rrez-Cuevas}
\email{rgutier2@ur.rochester.edu}
\affiliation{Center for Coherence and Quantum Optics, University of 
Rochester, Rochester, New York 14627, USA}
\affiliation{Institute of Optics, University of Rochester, Rochester, New 
York 14627, USA}
\author{Joseph H. Eberly}
\affiliation{Center for Coherence and Quantum Optics, University of 
Rochester, Rochester, New York 14627, USA}
\affiliation{Department of Physics and Astronomy, University of Rochester, 
Rochester, New York 14627, USA}

\date{\today}

\begin{abstract}
In the present manuscript, we present a high-speed method to control, 
manipulate and retrieve an intense vector soliton stored in the ground state 
coherences of a four-level atomic system. 
Additionally, we show the importance of the pulse area 
in determining the evolution of the system and present a constant in the 
evolution defined as the three-pulse area, a surprising extension to 
previously defined pulse areas.
\end{abstract}


\maketitle


\section{Introduction}

Ever since its discovery, the adiabatic technique of 
electromagnetically induced transparency (EIT) 
\cite{harris1997electromagnetically,*boller1991observation} has been the 
crux for light storage and manipulation. The interaction of a $\Lambda$ 
atomic system with a control field opens a transparency window for a resonant 
signal field to propagate through without absorption. As the group velocity 
is directly related to the properties of the optical susceptibility, this 
also gives a way to control the speed of propagation of the signal field all 
the way to where it can be stopped, and thus mapped into a spin wave. This 
procedure (and the retrieval of the stored signal) is best described in terms 
of dark-state polaritons (light-matter excitations) as was done in 
\cite{fleischhauer2000dark}.

This opened the door to numerous techniques for optical information storage 
where the main contenders are a far-off-resonant Raman scheme 
\cite{nunn2007mapping,*reim2011single} and a photon-echo based procedure 
\cite{moiseev2001complete}. For the first, the fields are highly detuned thus 
allowing the adiabatic elimination of the higher atomic level. 
Thus, the field 
is mapped into the ground states via stimulated Raman transitions. The second 
is an extension of the well-known phenomenon of photon echo in 
inhomogeneously broadened two-level atoms 
\cite{kurnit1964observation,*mossberg1982time}. A resonant signal photon 
wave-packet is absorbed and subsequentially mapped into the stable ground 
states by means of a short $\pi$-control pulse. The stored field can be 
recovered by a second counter-propagating  $\pi$-control pulse. Furthermore, 
in a series of papers 
\cite{gorshkov2007universal,*gorshkov2007photon,*gorshkov2007photon2,%
*gorshkov2007photon3,*gorshkov2008photon}, Gorshkov \emph{et al.}  brought 
all of these techniques into  a ``universal approach to optimal photon 
storage'' and devised a procedure to maximize the efficiency given any signal 
field.  

The enormous success of the $\Lambda$ configuration motivated the study of 
more complicated atomic systems, such as the double $\Lambda$ system where 
enhancement of nonlinear effects can be achieved as well as the storage of 
two signal pulses 
\cite{lukin2000resonant,*raczynski2004electromagnetically}. An $N$-type 
system has also been proposed for the control of two-photon absorption via 
quantum interference which can lead to an improvement in EIT 
\cite{harris1998photon,*yan2001nonlinear,*jiang2006optical}, as well as a  
giant enhancement of the Kerr nonlinearity \cite{niu2005giant}.

One of the most prolific extensions has been the 
four-level system in a tripodal configuration (see Fig.~\ref{fig:tripod}) 
where many advances have been made for light control and storage 
(e.~g.~double EIT  
\cite{paspalakis2002transparency,*wang2006large,*li2007two}). The formalism 
of dark-state polaritons \cite{fleischhauer2000dark} has been  extended for 
this four-level system in different kinds of scenarios. Depending on the 
initial preparation of the medium there can be either two signal pulses and 
one control, which leads to the possibility of storage of a photonic quantum 
bit \cite{karpa2008resonance}, or there can be just one signal pulse and two 
control pulses that allow two-channel-light storage 
\cite{raczynski2006polariton}. The existence of temporal and spatial vector 
solitons in this four-level system has also been shown 
\cite{hang2009weak,*liu2009ultraslow,*qi2011spatial}. 
Furthermore, it has been demonstrated that under the influence of a classical 
(intense) control field a propagating probe field acquires a phase that 
affects its state of polarization. There are proposals for using this effect 
to enhance the sensibility of Faraday magnetometers 
\cite{petrosyan2004magneto} or to serve as a polarization phase gate 
\cite{rebic2004polarization}.
In addition, some of these new phenomena 
have been demonstrated experimentally such as dark resonance switching 
\cite{ham2000coherence}, the existence of two transparency windows, enhanced 
crossed-phase modulation \cite{li2008enhanced}, the propagation of matched 
slow pulses \cite{macrae2008matched}, and two-field storage 
\cite{karpa2008resonance,wang2009slowing} (we refer the interested reader to 
these manuscripts for examples of experimental realizations of the tripodal 
scheme). 

The original work on pulse storage was tied to a requirement for adiabaticity 
(effectively near-constant intensity) of the 
control pulse which limits the speed of the process. This point was carefully 
studied by Matsko \emph{et al.}~\cite{matsko2001nonadiabatic}. They showed 
that the storage and retrieval are still possible by ``instantaneously'' 
switching the control field off and on. This was further studied by 
Shakhmuratov \emph{et al.}~\cite{shakhmuratov2007instantaneous} where they 
added the effect of an rf field in an $N$ configuration. Even though most 
proposals  work with cw control fields that are turned off and on, this might 
not be the best strategy. When one considers the optimization problem the 
resulting optimal field acquires a temporal structure  that clearly deviates 
from the standard cw field 
\cite{gorshkov2007universal,*gorshkov2007photon,*gorshkov2007photon2,%
*gorshkov2007photon3,*gorshkov2008photon,nunn2007mapping}. Another usual 
assumption is that the signal field is of quantum nature (low intensity). In 
\cite{dey2003storage} they go beyond these assumptions 
by means of a series of 
numerical experiments. But they restrict themselves to cw control fields and 
use the adiabatic theory of Grobe \emph{et al.}~\cite{grobe1994formation} to 
interpret their results as their signal pulses are long enough (the duration 
is about 100 times the lifetime in the excited state) that spontaneous 
emission needs to be included.

In the present work, we depart from the usual considerations for pulse 
storage as we consider the joint evolution of resonant-intense-broadband 
pulses (these are much shorter and about two orders of magnitude stronger 
that the ones considered in \cite{dey2003storage}). This is the realm of 
self-induced transparency (SIT), which was introduced by McCall and Hahn in 
their seminal papers \cite{mccall1967self,*mccall1969self}. They showed the 
crucial role that the total pulse area, defined as
\begin{equation}
\theta(x)=\int^{\infty}_{-\infty}\Omega(x,t)dt,
\label{eq:area}
\end{equation}
plays in this type of coherent interaction. Recent research has shown how 
the interaction of broadband pulses with matter opens the door to high-speed 
switching. In this framework, it has been shown that storage, manipulation 
and retrieval of a signal pulse in a $\Lambda$ system is possible 
\cite{groves2013jaynes} even in non-idealized conditions 
\cite{gutierrez2015manipulation} and how the methodology presented there can 
be extended to accommodate the storage of multiple pulses and added control 
of the information stored \cite{gutierrez2015multi}. A generalized two-pulse 
area was shown to play a role \cite{clader2007two}: $\Theta_{12}(x)=\sqrt{|\theta_1(x)|^2+|
\theta_2(x)|^2}$. 

Now, we extend this exploration to the tripodal atom 
interacting with three fields in resonance. We show how a vector soliton 
can be stored in the coherences of the ground states and then retrieved. We 
also find that the competing process of stealing population from the common 
excited state leads to a constraint on the area of each field as determined 
by a three-pulse area defined in Eq.~\eqref{eq:3area}. This is in 
alignment with previous results for the $\Lambda$ and double-$\Lambda$ 
systems \cite{clader2007two,groves2009multipulse}


\section{Theoretical model}

\begin{figure}
\centering
\includegraphics[scale=1]{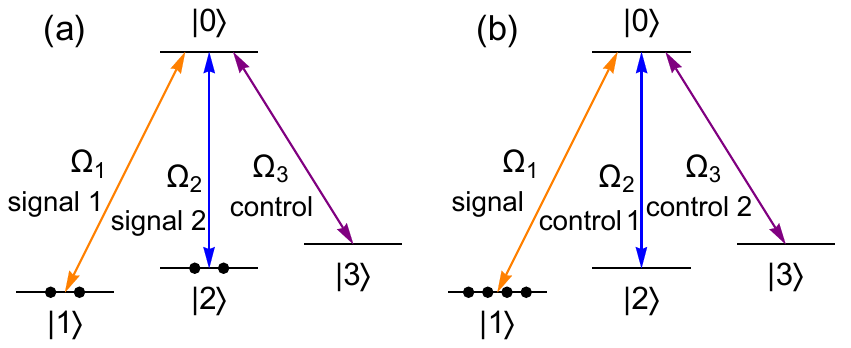}
\caption{\label{fig:tripod}(Color online) Four-level atom 
in a tripodal configuration,
interacting with three fields in resonance: (a) with the population 
initially distributed equally but incoherently among the ground states $
\ket 1$ and $\ket 2$ which leads to the presence of two signal pulses and 
one control and (b) with all the population in state $\ket 1$ which leads 
to one signal pulse and two control pulses.}
\end{figure}

For the tripodal atom (see Fig.~\ref{fig:tripod}), each ground state is only 
connected to the excited state 
$\ket 0$ by the dipole moment operator
\noindent and the fields are given by
\begin{equation}
\vec{E}(x,t)=\sum_{j=1}^3\vec{\mathcal E}_{j}(x,t)e^{i(k_{j0}x-\omega_{j0}
t)} +c.c.
\end{equation}
Here, $\omega_{10}$, $\omega_{20}$ and $\omega_{30}$ are the field 
frequencies, $k_{10}$, $k_{20}$ and $k_{30}$ are the vacuum wave numbers and 
$\vec{\mathcal E}_{1}(x, t )$, $\vec{\mathcal E}_{2}(x, t )$ and $
\vec{\mathcal E}_{3}(x,t)$ are the slowly-varying field envelopes. For 
simplicity the fields are taken to be at resonance and we will neglect the 
effects of Doppler broadening by considering a gas of cold atoms 
(given some minor substitutions, the results 
presented here will remain valid even in the presence of Doppler broadening 
\cite{gutierrez2016storage}). We will further assume that the 
fields are pulses short enough to neglect spontaneous emission but long 
enough so that the variation of the envelopes is much slower than the 
oscillations given by the optical frequency. This justifies the use of the 
slowly-varying-envelope approximation (SVEA). Considering the dipole and 
rotating-wave approximations, the Hamiltonian takes the form
\begin{equation}
 H=-\frac{\hbar}{2} \left( \Omega_1 \ket 0 \bra 1 + \Omega_2 \ket 0 \bra 2 +
 \Omega_3 \ket 0 \bra 3\right) + c.c.,
\label{hrwa}
\end{equation}
where we defined the Rabi frequencies $\Omega_{j}(x,t)=2\vec{d}_{0j}\cdot 
\vec{\mathcal E}_{j}(x, t )/\hbar$ and the non-zero off diagonal elements of $H$ can then be written as $H_{0j}=-\hbar \Omega_j(x,t)/2$ for $j=1,\,2,\,3$.

The dynamics of the field-matter system are described in terms of
the von Neumann equation for the density matrix,
\begin{equation}
i\hbar \pd{\rho}{t}=[ H, \rho],
\end{equation}
and Maxwell's wave equations in the slowly-varying envelope approximation
\begin{subequations}
\label{meqs}
\begin{align}
\left(\pd{ }{x}+\frac{1}{c}\pd{}{t}\right)\Omega_{1}&=i\mu_{10} \rho_{01},\\
\left(\pd{ }{x}+\frac{1}{c}\pd{}{t}\right)\Omega_{2}&=i\mu_{20} \rho_{02},\\
\left(\pd{ }{x}+\frac{1}{c}\pd{}{t}\right)\Omega_{3}&=i\mu_{30} \rho_{03}.
\end{align}
\end{subequations}
where we defined the atom-field coupling parameters as 
$\mu_{j0}=N\omega_{j0}|d_{j0}|^2/
\hbar \epsilon_0c$ with $j=1,\,2,\,3$.
These give a set 
of nonlinear partial differential equations that need to be solved 
simultaneously. Therefore, we cannot move forward, at least analytically, 
without making further assumptions. The key lies in considering equal 
atom-field coupling parameters for all three transitions, $\mu_{10}=\mu_{20}=
\mu_{30}=\mu$ (this assumption might not be realistic but given the 
stability of this type of solution in non-idealized conditions shown in 
\cite{gutierrez2015manipulation} we expect the results to remain valid for 
the most part). Doing so, and 
introducing the constant matrix $W=i\ket 0 \bra 0$, we can write the 
evolution equations as
\begin{equation}
\label{eq:mb}
i\hbar \pd{\rho}{T}=[ H, \rho], \quad \text{and} \quad \pd{H}{Z}=-
\frac{\hbar \mu}{2}[W, \rho],
\end{equation}
in terms of the traveling-wave coordinates $T=t-x/c$ and $Z=x$.
By computing the commutator, one can verify that the matrix 
equation for the field [Eqs.~\eqref{eq:mb}] 
indeed reduces to Eqs.~\eqref{meqs} plus some (irrelevant) trivial equations.

From the form in which the evolution equations were written 
in Eqs.~(\ref{eq:mb}) it is easy to show that the system is integrable 
(Appendix \ref{app:meth}) and therefore solvable by 
standard methods such as inverse scattering 
\cite{gardner1967method,*ablowitz1973nonlinear,*lamb1980elements,
*chakravarty2014inverse},  the 
B\"acklund transformation 
\cite{lamb1971analytical,*miura1976backlund,*park1998field} and 
the Darboux transformation \cite{gu2006darboux,cieslinski2009algebraic}, to 
name a few. This study contrasts with many previous results in that it 
considers the simultaneous evolution of the three fields instead of assuming
one of them to be a strong constant field that induces some extra 
nonlinearities in the evolution of the other two 
\cite{petrosyan2004magneto,rebic2004polarization,hang2009weak,%
*liu2009ultraslow,*qi2011spatial}. In addition, we consider the full nonlinear 
interaction which contrasts to the first-order approximation in the fields 
done in \cite{paspalakis2002transparency}.


\section{Two-pulse storage}

We now proceed to solve the Maxwell-Bloch 
equations for a tripodal atomic system [Eqs.~(\ref{eq:mb})] using the single-
soliton Darboux transformation and the nonlinear superposition rule. This 
method allows us to start from a trivial solution and obtain complicated 
pulse dynamics by some integration and algebraic manipulation. A 
review of this method can be found in Refs.~
\cite{cieslinski2009algebraic,gutierrez2015multi} and is similar to the one 
presented by Clader and Eberly in \cite{clader2007two}. For the interested 
reader, an outline of the principal steps is presented 
in Appendix \ref{app:meth}.
 
\begin{figure}
\centering
\includegraphics[scale=1.]{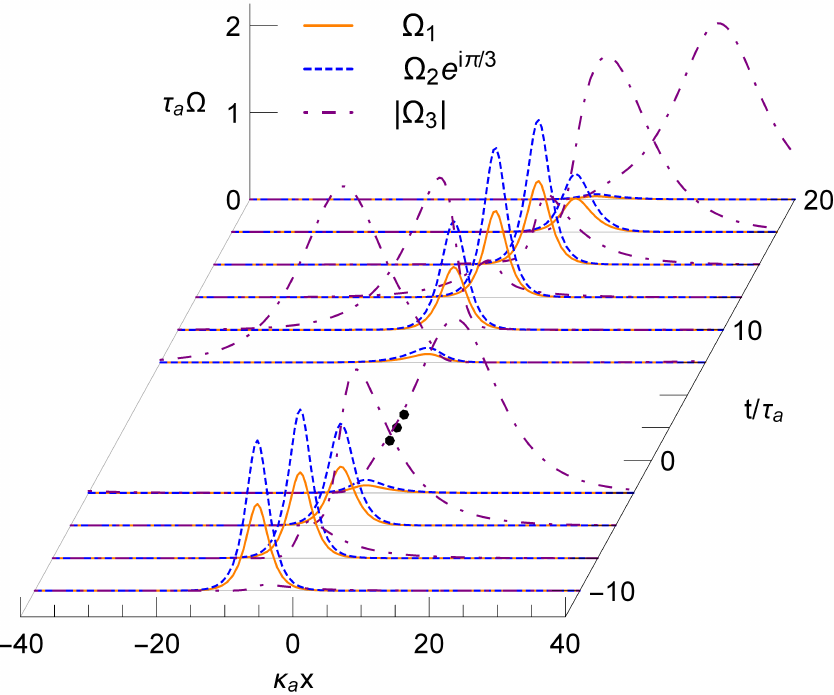}
\caption{\label{fig:pulses1}(Color online) Pulse 
evolution dictated by the second order 
solution obtained from the medium preparation shown in Fig.~\ref{fig:tripod}
(a). The encoding of the vector soliton is separated by the ellipses from 
its retrieval and displacement.}
\end{figure}

We will start by taking the situation depicted in Fig.~\ref{fig:tripod}(a) 
as our trivial solution, that is, an incoherent preparation of the medium 
given by $\rho=1/2(\ket 1 \bra 1 + \ket 2 \bra 2)$ and no fields. Applying 
the Darboux transformation to this seed solution, we obtain a first-order 
solution for which the analytic expressions can be simplified in the limits 
of infinitely long negative and positive times (the expression of the 
involution matrix, used to compute the solution in these limits is 
presented in Table I of Appendix \ref{app:lim}). This provides 
us with the state of the system before and after the interaction. An example 
of the pulse dynamics is represented in Fig.~\ref{fig:pulses1}.

Initially ($T/\tau_a\ll -1$), we have two SIT pulses of duration $\tau_a$ 
propagating throughout the medium at a reduced group velocity for the two 
signal fields. Their shapes are given by
\begin{subequations}\label{eq:om1}
\begin{equation}
\Omega_1=\frac{2}{\tau_a}\cos \left( \frac{\nu}{2}\right)\sech 
\left(\frac{T}{\tau_a}-\frac{\kappa_a}{4} Z+\eta \right),
\end{equation}
\begin{equation}
\Omega_2=\frac{2}{\tau_a}e^{i\phi}\sin \left(\frac{\nu}{2}\right)\sech 
\left(\frac{T}{\tau_a}-\frac{\kappa_a}{4} Z+\eta \right),
\end{equation}
\end{subequations}
where we introduced the absorption coefficient $\kappa_a=\mu \tau_a/2$ and 
the constants of integration $\eta$, $\phi$ and $\nu$. The angles $\phi$ and 
$\nu$ define the area of each signal pulse, $\theta_1=2\pi\cos \left(\nu/
2\right) $ and $\theta_2=2\pi e^{i\phi}\sin \left(\nu/2\right)$ (the area of 
the first pulse was taken to be real in order to fix the global phase). Note 
that the signal pulses are ``normalized'' by the total two-pulse area (as 
defined by Clader and Eberly in \citep{clader2007two}), that is, $\sqrt{|
\theta_1|^2+|\theta_2|^2}=2\pi$. In this limit, the control pulse tends 
towards zero. If we assume that the two fields address different transitions 
due to their polarization as in 
\cite{hang2009weak,*liu2009ultraslow,*qi2011spatial}, then the stored field 
can be written as
\begin{align} \label{eq:vect}
\vec{E}_s(x,t)=&\frac{\hbar}{d \tau_a}\left[\cos \left(\frac{\nu}{2}\right) 
\vec{p_1}+e^{i\phi}\sin \left(\frac{\nu}{2}\right)\vec{p_2}\right] \nonumber \\
&\times\sech \left(\frac{T}{\tau_a}-\frac{\kappa_a}{4} Z+\eta 
\right)e^{i(k_{s}x-\omega_{s}t)} +c.c.
\end{align}
to show that it can be seen as a vector soliton ($\vec{p_1}$ and $\vec{p_2}$ 
are two orthogonal polarization vectors). Given the free parameters $\phi$ 
and $\nu$ the polarization can occupy any point on the surface of the 
Poincar\'e sphere.

\begin{figure}
\centering
\includegraphics[width=1.\linewidth]{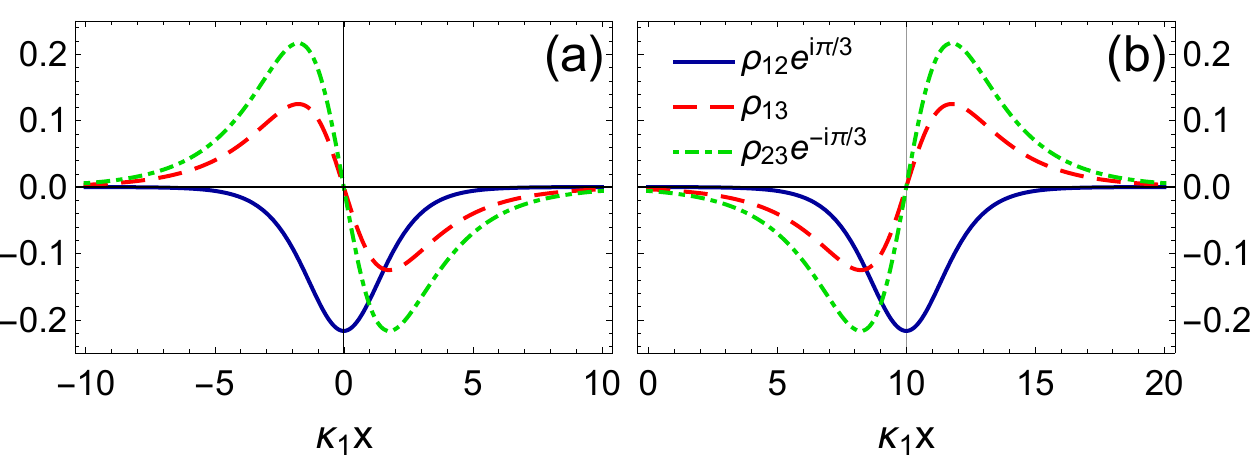}
\caption{\label{fig:den1}(Color online) Imprint as 
it has been encoded by the pulse 
dynamics shown in Fig.~\ref{fig:pulses1} in the ground state coherences 
(a) before and (b) after the displacement.}
\end{figure}

As the signal pulses propagate, the control pulse starts to drive some of 
the population to the ground state $\ket 3$ thus amplifying its 
asymptotically small amplitude while depleting the signal pulses. This 
transfer slowly takes over until we are left with a decoupled control pulse 
propagating away with the light's phase velocity. The full solution for the 
pulses is provided in Eqs.~\eqref{app:comvec}. During the interaction, 
the information of the vector soliton is imprinted into the ground state 
elements of the density matrix in the form of a spin-wave. This ``imprint''
is depicted in 
Fig.~\ref{fig:den1}. Even if the coherences are non-zero everywhere 
(because of the infinitely long tails of the sech function) it is clear 
that they are well localized around their center which we identify as 
their location. Computing the total pulse area for each pulse 
[see Eqs.~\eqref{app:areavec}] we get
\begin{equation}
\label{eq:imp1}
\frac{\sqrt{|\theta_1(x)|^2+|\theta_2(x)|^2}}{|\theta_3(x)|}=e^{-
\frac{\kappa_a}{2} (x-x_1)},
\end{equation}
where $x_1$ gives the location of the imprint. Therefore, the imprint is 
located at the position where the two-signal-pulse area is equal to the 
control pulse area.

This two-pulse storage, when taken to low intensities (quantum states of 
light), can be seen as a qubit storage procedure as the phase between the 
two pulses is also encoded within the atomic medium. This was actually done  
in \cite{karpa2008resonance} but was based on the usual EIT procedure of 
slowing down the quantum signal by means of a classical control pulse.


Taking a closer look to the expressions for the area of each pulse in the 
first-order solution, we see that a role for a three-pulse area is implied:
\begin{equation}
\label{eq:3area}
\Theta_{123}(x)=\sqrt{|\theta_1(x)|^2+|\theta_2(x)|^2+|\theta_3(x)|^2}=2\pi.
\end{equation}
This result is a clear statement of the relationship between this solution 
and the original SIT solution. The pulse area is a key quantity, not only 
controlling the reshaping of pulses but also the storage and manipulation of 
the information stored in the medium. Its importance goes well beyond the 
two-level atom remaining a constant in pulse dynamics for multi-level 
systems \cite{clader2007two,groves2009multipulse}. For non-idealized input 
pulse shapes, these multi-level systems follow a similar behavior to the 
predictions of the area theorem for  a two-level system 
\cite{clader2007two,clader2008two,groves2009multipulse} even in the 
absence of  Doppler broadening \cite{gutierrez2015manipulation}. 

Other results referring to pulse area theorems have been worked out for the $
\Lambda$ configuration. The first was done by Tan-no 
\emph{et al.}~\cite{tan1972two} but is 
limitied to highly-detuned fields in two-photon 
resonance. Clader and Eberly presented a clear comparison of this stimulated 
Raman scattering (SRS) with the exact Maxwell-Bloch equations 
\cite{clader2007two}. Another famous result is the dark area theorem 
\cite{eberly2002wave}, where the dark area is not 
defined in the usual manner (it 
has time derivatives of the Rabi frequencies). Nevertheless, it provides a 
simple spatial evolution of a quantity involving both Rabbi frequencies which 
helps understand the interaction of the two fields.  This result is very much 
in line with the original area theorem \cite{mccall1967self,*mccall1969self}. 
A more recent result was presented by Shchedrin \emph{et al.}~in 
\cite{shchedrin2015analytic}. There they consider pulse interaction beyond 
the rotating wave approximation extending the applicability of their results. 
However, the pulse area theorem lacks any mention of spatial evolution and is 
in fact formulated more like a energy conservation equation which still 
involves the Rabi frequencies. To this we could add the conservation laws 
that can be deduced from the integrability of the Maxwell-Bloch equations 
(see for example \cite{chakravarty2015soliton}). As we already mentioned our 
result is similar to the ones presented in 
\cite{clader2007two,groves2009multipulse} and thus we expect homologous  
results for the case of partially mixed states \cite{clader2008two}.


Another valid solution, that we can obtain from the same seed solution (by 
an appropriate choice of integration constants, see Appendix \ref{app:lim}), 
is that of a 
sech-shaped-$2\pi$-control pulse propagating through the medium at the speed 
of light in vacuum. Superimposing this solution with the one previously 
described, we obtain a second-order solution that (by an appropriate choice 
of parameters) describes a well-defined pulse sequence. Here, we first have 
the storage of the signal pulses and then the collision of another control 
pulse, of different duration $\tau_b$, with the imprint. Upon interaction 
with the imprint, the control pulse retrieves the stored signal pulses along 
with the pulse area and relative phase information. The retrieved pulses 
then propagate farther into the medium and then are again stored, but in a 
displaced location (see Figs.~\ref{fig:pulses1} and \ref{fig:den1}). The 
displacement of the imprint is controlled by the duration of the pulses via 
\begin{equation}
\label{eq:dis1}
\delta =\kappa_a(x_2-x_1)=2\ln \left| \frac{\tau_a+\tau_b}{\tau_a-\tau_b}
\right|,
\end{equation}
where $x_2$ is the new location of the imprint.  

In reality, we are not going to have an infinite medium, but this can be 
used to our advantage. Given a finite medium, Eq.~\eqref{eq:imp1} tells us 
how we should map this solution to initial conditions as the ratio of the 
areas of the entering pulses will control the location of the imprint. Now, 
if the duration $\tau_b$ is tailored so that the displacement given by 
Eq.~\eqref{eq:dis1} is larger than the 
length of the medium, the control pulse 
will move the imprint outside. This would frustrate the re-encoding of the 
signal pulses and thus effectively retrieve the information stored in the 
atomic medium. Therefore, the finiteness of the medium provides us with the 
means to retrieve the stored pulses. Additionally, there could be some 
residual coherence between the ground states $\ket 1$ and $\ket 2$ from the 
initial preparation. But, using the same formalism, we can show that the 
storage-retrieval procedure is still viable  (an extended discussion about 
the effects of partial coherence in a $\Lambda$ system is presented in 
\cite{clader2008two} and one would expect similar results for this case).



\section{Two-channel memory}
\label{sec:tcm}

\begin{figure}
\centering
\includegraphics[scale=1.]{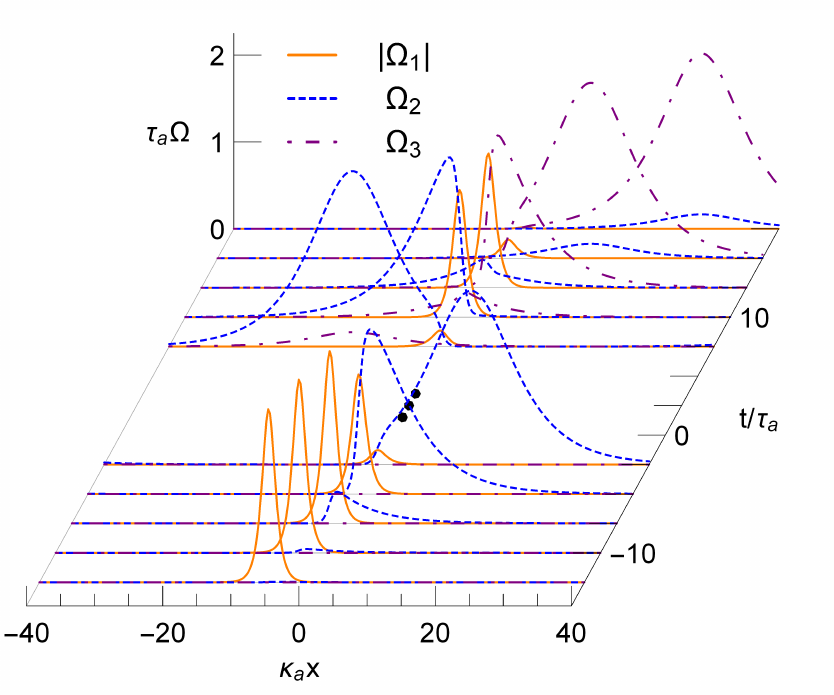}
\caption{\label{fig:pulses2}(Color online) Pulse 
evolution dictated by the second order 
solution obtained from the medium preparation shown in Fig.~\ref{fig:tripod}
(b). The encoding of the signal pulse is separated by the ellipses from the 
channel-switching and displacement.}
\end{figure}

The three-pulse area is, of 
course, not limited to the specific solution of the storage of a vector 
soliton. We can consider another possibility for the seed solution, such as 
the one depicted in Fig.~\ref{fig:tripod}(b). The medium is prepared in the 
ground state $\ket 1$  and again with no fields. In this case, 
the first-order solution describes a signal 
pulse that is stored via the interaction 
of two control pulses with the atomic system as intermediary. The pulse 
dynamics start with an SIT signal pulse of duration $\tau_a$ propagating at 
a reduced velocity. As this pulse propagates, the asymptotically small 
control pulses start to drive some of the population to the ground states $
\ket 2$ and $\ket 3$ thus amplifying them while depleting the signal pulse.
The full solution for the 
pulses is provided in Eqs.~\eqref{app:comch}. 
During this interchange, the signal pulse is encoded in the coherences of 
the ground states. We can think of this system as having two channels, 
represented by the coherences $\rho_{12}$ (channel 1-2) and $\rho_{13}$ 
(channel 1-3), connected by the coherence $\rho_{23}$. Computing the total 
pulse area for each pulse [see Eqs.~\eqref{app:areach}] we get the 
following relation,
\begin{equation} \label{eq:cimp1}
\frac{|\theta_1(x)|}{\sqrt{|\theta_2(x)|^2+|\theta_3(x)|^2}}=e^{-\kappa_a(x-
x_1)}
\end{equation}
where $x_1$ denotes the location of the imprint. In a similar fashion to the 
previous solution, the initial location of the imprint is determined by the 
ratio between the signal pulse area and the two control pulse areas. The 
amount of information stored in channel 1-2 is determined by the ratio 
between the area of the control pulse $\Omega_2$
and the two-control-pulse area squared, 
$r_{1-2}=|\theta_2(x)|^2/(|\theta_2(x)|^2+|\theta_3(x)|^2)$, and similarly 
for channel 1-3, $r_{1-3}=|\theta_3(x)|^2/(|\theta_2(x)|^2+|\theta_3(x)|^2)
$. The particular case of storing in just one channel can be seen in 
Fig.~\ref{fig:pulses2}. The possibility for two-channel storage had already 
been mentioned in \cite{raczynski2006polariton} but this study was based on 
an EIT type interaction.

Here again, by studying the expressions for the individual pulse areas, we 
find that by summing them as prescribed by Eq.~\eqref{eq:3area} we obtain $
\Theta_{123} (x)=2\pi$. This goes to show that the three-pulse area is not 
limited to a specific type of preparation for the system. Now, the solution 
just presented is interesting by itself. Therefore, its interaction with 
subsequent pulses is well deserving of mention.  

\begin{figure}
\centering
\includegraphics[scale=1]{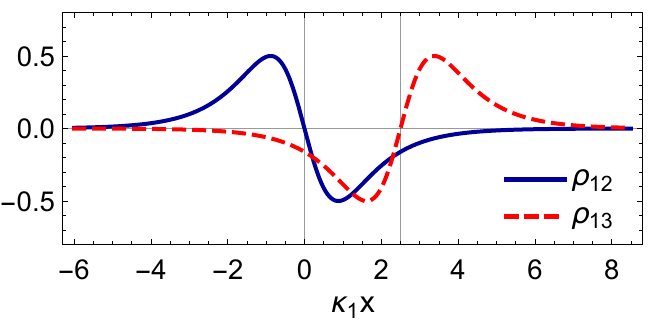}
\caption{\label{fig:den2}(Color online) Imprint 
as it has been encoded by the pulse 
dynamics shown in Fig.~\ref{fig:pulses2} in the ground state coherences 
before (continuous lines) and after (dashed lines) the channel-switching.}
\end{figure}

Let's consider the second-order solution born out of the superposition of 
the solution previously discussed and that of two control pulses of total 
two-pulse area equal to $2\pi$ and duration $\tau_b$ decoupled from the 
medium, we discover that the information can be displaced between channels 
by the subsequent interaction of the imprint with other control pulses. For 
simplicity, we will consider the case in which the imprint was only 
encoded into channel 1-2 and we want to move all the information into the 
other channel (this case is depicted in Figs.~\ref{fig:pulses2} and 
\ref{fig:den2}). In this scenario, simple expressions can be worked out for 
the necessary pulse areas to achieve the channel-switching of the 
information encoded in the medium and its displacement. These are given by
\begin{equation}
|\theta_2|=2\pi \sqrt{\frac{1+\tau_b/\tau_a}{2}} \quad \text{and} \quad |
\theta_3|=2\pi \sqrt{\frac{1-\tau_b/\tau_a}{2}}
\end{equation}
and the corresponding displacement of the imprint is
\begin{equation}
\delta=\frac{1}{2}\ln\left(\frac{\tau_a+\tau_b}{\tau_a-\tau_b}\right).
\end{equation}
We note that the two things are related and depending on the duration of the 
control pulses we will have to choose the appropriate pulse area. Here, we 
assumed that $\tau_a>\tau_b$. The relative phase between the two control 
pulses determines the phase of the coherence which in turn dictates the 
phase of the retrieved pulse. The channel-switching of the imprint is shown 
in Fig.~\ref{fig:den2}. Additionally, as long as the signal pulse is only 
stored in one channel, we can displace the imprint and retrieve it (if we 
consider a finite medium) by the results already obtained for a $\Lambda$ 
system \cite{groves2013jaynes,gutierrez2015manipulation}.

It is also interesting to note that, together with the displacement of the 
imprint, the intensity of the control pulses is inverted (see 
Fig.~\ref{fig:pulses2}). If the two control pulses address different 
transitions due to their polarization then this inversion of intensity is 
actually a rotation in the polarization state of the vector-control pulse. 
This change in the polarization in the tripodal 
configuration has already been studied in 
different regimes of light \cite{petrosyan2004magneto,rebic2004polarization}. 
Of course the change of polarization is directly tied to the initial imprint, 
namely its phase and the ratios $r_{1-2}$ and $r_{1-3}$. 

It is also relevant 
to mention that if we had considered an initial preparation of the 
medium analogous to the one in \cite{byrne2003polarization} (i.~e.~incoherent 
superposition of the ground states), we would have obtained 
a three-color switching 
analogous to the polarization switching mentioned in their work along 
with similar stochastic dynamics when dealing with disordered populations 
\cite{atkins2012stochastic,*newhall2013random}.

\section{Summary}

In summary, we have presented analytic solutions to 
the Maxwell-Bloch equations for a tripodal system which suggest the 
possibility of high-speed storage and retrieval of a vectorial soliton as 
well as that of a two-channel memory. These results are similar to the ones 
derived in \cite{raczynski2006polariton,karpa2008resonance} that are based 
on the dark-state polariton formalism which inevitably carries with it the 
adiabatic approximation limiting the speed of the processes involved. 
Still, it is interesting to note the similarities of our results to 
previous treatments dealing with quantum states of light entailing 
the free space propagation of the control pulse and thus the omission
of any change of shape during propagation.
We also defined an extension for the area theorem in this three-pulse 
scenario 
which is not tied to a specific preparation of the medium. This result 
raises new questions, such as: What is the appropriate way to combine the 
individual pulse areas in given multi-level and multi-pulse systems? 
Clearly, there must be a dependence on the connection between atomic levels 
and the number of ground- and excited-states. We expect these results to be 
of scientific importance for the continuing development of the pulse 
storage and manipulation field.

\section*{Acknowledgments}

This work was supported by NSF through Grant nos. (PHY-1203931, PHY-1505189) 
and a CONACYT fellowship awarded to R.~G.~C.

 \appendix

 \section{Darboux transformation and nonlinear superposition rule }
 \label{app:meth}

{
\renewcommand{\arraystretch}{1.7}
\setlength{\tabcolsep}{15pt}
\begin{table*}
\caption{\label{tab:Ma1} Elements of the involution matrix $M^a$ for the 
two-pulse storage first-order solution showing only the 
elements in the limits of infinite negative and positive times as these can 
be written in a simple form. 
\footnote{We defined the parameters $\bar T=T/\tau_a$, $\bar Z=\mu \tau_aZ/
      2$, $|c_{+}|^2=|c_1|^2+ |c_2|^2$, $\eta_{jk}=\ln |c_j/c_k|$. }}
\centering
\begin{tabular}{ccc}
\hline
\hline
& $T/\tau_a \ll-1$ & $T/\tau_a \gg1$  \\
 \hline
$M_{00}^a$&$-\tanh \left(\bar T -\frac{\bar Z}{2}+\eta_{+0}\right)$&$-1$\\
$M_{11}^a$&$\frac{|c_1|^2}{|c_+|^2}\left[1+\tanh \left(\bar T -\frac{\bar Z}
{2}+\eta_{+0}\right)\right]-1$&$\frac{|c_1|^2}{|c_+|^2}\left[1-\tanh \left( -
\frac{\bar Z}{2}+\eta_{+3}\right)\right]-1$ \\
$M_{22}^a$&$\frac{|c_2|^2}{|c_+|^2}\left[1+\tanh \left(\bar T -\frac{\bar Z}
{2}+\eta_{+0}\right)\right]-1$&$\frac{|c_2|^2}{|c_+|^2}\left[1-\tanh \left( -
\frac{\bar Z}{2}+\eta_{+3}\right)\right]-1$\\
$M_{33}^a$&$-1$&$-\tanh \left( -\frac{\bar Z}{2}+\eta_{+3}\right)$ \\
$M_{01}$&$ \frac{c_0c_1^*}{|c_0c_+|}\sech\left(\bar T -\frac{\bar Z}{2}+
\eta_{+0} \right) $&$0$ \\
$M_{02}$&$ \frac{c_0c_2^*}{|c_0c_+|}\sech\left( \bar T -\frac{\bar Z}{2}+
\eta_{+0} \right) $&$0$\\
$M_{03}$&$0$&$\frac{c_0c_3^*}{|c_0c_+|}\sech\left( \bar T +\eta_{30} \right)$ 
\\
$M_{12}^a$&$0$& $ \frac{c_1c_2^*}{|c_+|^2}\left[1+\tanh\left( -\frac{\bar Z}
{2}+\eta_{+3} \right) \right]$ \\
$M_{13}^a$&$0$&$\frac{c_1c_3^*}{|c_3c_+|}\sech\left( -\frac{\bar Z}{2}+
\eta_{+3} \right)$   \\
$M_{23}^a$ &$0$&$\frac{c_2c_3^*}{|c_3c_+|}\sech\left( -\frac{\bar Z}{2}+
\eta_{+3} \right)$  
\\
\hline
\hline
\end{tabular}
\end{table*}
}

In what follows we will present an outline of the steps to take in order to 
obtain the desired solution. The interested reader is referred to more 
complete treatments, such as those found in 
\cite{gutierrez2015multi,*cieslinski2009algebraic}. 
We start by defining the operators $U(\lambda)=-(i/\hbar) H - 
\lambda W$ and $ V(\lambda)=(i\mu/2\lambda) \rho$, where $\lambda$ is a 
constant known 
as the spectral parameter. Using the Maxwell-Bloch equations [Eqs.~(4) in the 
main text] it is easy to show that the equation
\begin{equation}
\partial_Z  U-\partial_T V +[U,V]=0
\end{equation}
is satisfied. This is known as the Lax equation and is used as a criterion 
for the integrability of a system of equations. The operators $U$ and $V$ are 
called the Lax operators. Therefore, it is possible to obtain solutions by 
methods such as the Darboux transformation. 

The idea behind this method is to build complicated (and interesting) 
solutions out of trivial ones. Hence, the first step is to find a trivial  
solution for our system of equations. An example is the one used for the 
two-channel-memory solution where the fields are taken equal to zero and the 
medium prepared in the ground state $\ket 1$. This trivial or zeroth order 
solution is identified the density matrix $\rho^0$ and the Hamiltonian $H^0$ 
which give the corresponding Lax operators $U^0$ and $V^0$. The next step is 
to solve the following linear equations
\begin{subequations}
\begin{align}
\left(I \partial_T-U^0(-\lambda_a)\right)\ket \varphi =0, \\
\left(I \partial_Z-V^0(-\lambda_a)\right)\ket \varphi =0.
\end{align}
\label{eqphi}
\end{subequations}
These equations are derived by requiring the new Lax pair to satisfy the 
Lax equation which is equivalent to the Maxwell-Bloch equations and demanding 
a number of additional properties, such as conservation of the spectral 
dependence and hermiticity of the density matrix and Hamiltonian.
By solving them we obtain the vector $\ket \varphi$ in terms of which a 
unitary involution is defined as
\begin{equation}
M^a=2\frac{\ket \varphi \bra \varphi}{\braket{\varphi}{\varphi}}-I.
\end{equation}
Finally, from this matrix we can construct a new first-order solution using
\begin{subequations}
\begin{equation}
H^a=H^{0}-i\hbar \lambda_a\left[ M^a,W \right],
\end{equation}
\begin{equation}
\rho^a=M^a \rho^0 M^a.\label{mden}
\end{equation}
\end{subequations}

In principle, the procedure just described could be used to 
compute higher-order solutions, but the reality is that Eqs.~(\ref{eqphi}) 
become harder to solve with each step. Fortunately, there is no need due to 
the existence of a nonlinear superposition rule. 
Given a seed (zeroth-order) solution we can derive a first-order solution
with parameter $\lambda_a$ and Lax operators $U^a$ and $V^a$, 
as just described. From this solution, another second-order solution can 
be constructed by solving Eqs.~\eqref{eqphi} with parameter $\lambda_b$ 
and Lax operators $U^{b}$ and $V^{b}$ and thus obtain the second-order
involution $M^{ab}$ and Lax operators $U^{ab}$ and $V^{ab}$. 
Alternatively, we could have  first computed the
first-order solution with parameter $\lambda_b$ and from this 
obtained the second-order solution identified by Lax operators 
$U^{ba}$ and $V^{ba}$. The theorem of permutability states that there is
nothing special about the order in which we compute the solutions, thus both 
second-order solutions 
should be the same, that is $U^{ab}=U^{ba}$ and $V^{ab}=V^{ba}$. 
From this condition we obtain the nonlinear superposition rule
which gives us the involution matrix
\begin{equation}
M^{ab}=(\lambda_a M^a-\lambda_b M^b)(\lambda_a M^a M^b -\lambda_b I)^{-1},
\label{nls}
\end{equation}
and the second-order density matrix and Hamiltonian are
\begin{subequations}
\begin{equation}
\label{denab}
\rho^{ab}=M^{ab}M^a\rho^0 M^a M^{ab}
\end{equation}
and
\begin{equation}
H^{ab}=H^0-i \hbar 
(\lambda_a^2-\lambda_b^2)\left[(\lambda_aM^a-\lambda_bM^b),W\right].
\end{equation}
\end{subequations}

 \section{First-order solutions and limiting expressions}
 \label{app:lim}

\subsection{Two-pulse storage}

{
\renewcommand{\arraystretch}{1.7}
\setlength{\tabcolsep}{24pt}
\begin{table*}
\caption{\label{tab:Ma2} Elements of the involution matrix $M^a$ for the 
two-channel memory first-order solution showing only the 
elements in the limits of infinite negative and positive times as these can 
be written in a simple form.
\footnote{ We defined the parameters $\bar T=T/\tau_a$, $\bar Z=\mu \tau_aZ/
      2$, $|d_{+}|^2=|d_2|^2+ |d_3|^2$, $\xi_{jk}=\ln |d_j/d_k|$. } }
\centering
\begin{tabular}{ccc}
\hline
\hline
& $T/\tau_a \ll-1$ & $T/\tau_a \gg1$ \\
 \hline
$M_{00}^a$&$-\tanh \left(\bar T -\bar Z+\xi_{10}\right)$ & $-1$ \\
$M_{11}^a$&$\tanh \left(\bar T -\bar Z+\xi_{10}\right)$ & $\tanh \left( -\bar 
Z+\xi_{1+}\right)$ \\
$M_{22}^a$& $-1$ &$\frac{|d_2|^2}{|d_+|^2}\left[1-\tanh \left( -\bar Z+
\xi_{1+}\right)\right]-1$\\
$M_{33}^a$&$-1$ & $\frac{|d_3|^2}{|d_+|^2}\left[1-\tanh \left( -\bar Z+
\xi_{1+}\right)\right]-1$ \\
$M_{01}$& $ \frac{d_0d_1^*}{|d_0d_1|}\sech\left( \bar T -\bar Z+\xi_{10} 
\right) $ &$0$ \\
$M_{02}$&$0$ &$\frac{d_0d_2^*}{|d_0d_+|}\sech\left( \bar T+\xi_{+0} \right)$ 
\\
$M_{03}$& $0$ &$\frac{d_0d_3^*}{|d_0d_+|}\sech\left( \bar T +\xi_{+0} \right)
$ \\
$M_{12}^a$&$0$&$\frac{d_1d_2^*}{|d_1d_+|}\sech\left( -\bar Z+\xi_{1+} \right) 
$ \\
$M_{13}^a$&$0$ & $\frac{d_1d_3^*}{|d_1d_+|}\sech\left( -\bar Z+\xi_{1+} 
\right)$  \\
$M_{23}^a$ &$0$& $ \frac{d_2d_3^*}{|d_+|^2}\left[1-\tanh\left( -\bar Z+
\xi_{1+} \right) \right]$ \\
\hline
\hline
\end{tabular}
\end{table*}
}

Taking the seed to be the trivial solution of a medium in an incoherent 
superposition of the ground states $\ket 1$ and $\ket 2$ ($\rho^0=1/2(\ket 1 
\bra 1 + \ket 2 \bra 2)$) and fields off ($H^0=0$), Eqs.~\eqref{eqphi} can be 
solved to obtain
\begin{equation}
\ket{\varphi^a}=
\left(\begin{array}{c}
c_0 e^{-T/\tau_a}\\
c_1 e^{-\mu \tau_a Z/4}\\
c_2  e^{-\mu \tau_a Z/4}\\
c_3
\end{array}
\right).
\end{equation}
Here, the spectral parameter was written as $\lambda_a=i/\tau_a$, with 
$\tau_a \in \mathbb{R}$ and where $c_0$, $c_1$, $c_2$, $c_3$ are constants of 
integration. From this it is easy to compute the $M$ matrix but the exact 
expression of its elements is cumbersome and not very illustrative. 
Therefore, we limit ourselves to write the simplified expressions for large 
positive and negative times in Table~\ref{tab:Ma1}. 

From the constants of integration we can define other, more representative 
parameters:
\begin{subequations}
\begin{equation}
\cos \left( \frac{\nu}{2}\right)= i \frac{c_0 c_1^*}{|c_0 c_+|},
\end{equation}
\begin{equation}
e^{i\phi}\sin \left( \frac{\nu}{2}\right)= i \frac{c_0 c_2^*}{|c_0 c_+|},
\end{equation}
\begin{equation}
\eta=\eta_{+0}=\ln \left| \frac{c_+}{c_0}\right|,
\end{equation}
\begin{equation}
\kappa_a x_1= 2 \eta_{+3}=2 \ln \left| \frac{c_+}{c_3}\right|.
\end{equation}
\end{subequations}
These are the parameters used in Eqs.~(5-7). The most general solution (when 
all integration constants are not zero) gives us the solution that describes 
the storage of the vector soliton. This solution is then superimposed with 
another solution describing a control pulse propagating by itself decoupled 
from the medium. This second solution is obtained from setting $c_1=c_2=0$. 
 
For the first-order solution, the complete form for the 
fields is given by
\begin{subequations}
\label{app:comvec}
\begin{align}
\Omega_1&=\frac{4 i c_0 c_1^*}{\tau_a} e^{-\kappa_a Z/2 }f(Z,T),\\
\Omega_2&=\frac{4 i c_0 c_2^*}{\tau_a} e^{-\kappa_a Z/2 }f(Z,T),\\
\Omega_3&=\frac{4 i c_0 c_3^*}{\tau_a} f(Z,T),
\end{align}
with
\begin{equation}
f(Z,T)=\frac{e^{-T/\tau_a}}{|c_0|^2 e^{-2T/\tau_a}+|c_+|^2
e^{-\kappa_a Z}+|c_3|^2}.
\end{equation}
\end{subequations}
It is clear that the integration with respect $T$ for the pulse areas 
only involves the function $f$ and
\begin{equation}
\int^{\infty}_{-\infty}f(Z,T) dT= \frac{\tau_a \pi}{2|c_0|}\left(
 |c_+|^2e^{-\kappa_a Z}+|c_3|^2 \right)^{-1/2}.
\end{equation}
The pulse area can be readily written as
\begin{subequations}
\label{app:areavec}
\begin{align}
\theta_1&=\frac{2\pi i c_0 c_1^*}{|c_0|}\frac{ e^{-\kappa_a Z/2 } }
{\sqrt{ |c_+|^2e^{-\kappa_a Z}+|c_3|^2 }},\\
\theta_2&=\frac{2\pi i c_0 c_2^*}{|c_0|}\frac{ e^{-\kappa_a Z/2 } }
{\sqrt{ |c_+|^2e^{-\kappa_a Z}+|c_3|^2 }},\\
\theta_3&=\frac{2\pi i c_0 c_3^*}{|c_0|}\frac{ 1 }
{\sqrt{ |c_+|^2e^{-\kappa_a Z}+|c_3|^2 }}.
\end{align}
\end{subequations}
The results given in Eqs.~\eqref{eq:imp1} and \eqref{eq:3area} follow.

\subsection{Two-channel memory}

Now let us consider the other seed solution considered in the main text, the 
medium in the ground state $\ket 1$ ($\rho^0=\ket 1 \bra 1 $) and fields off 
($H^0=0$). We solve Eqs.~\eqref{eqphi} and obtain
\begin{equation}
\ket{\varphi^a}=
\left(\begin{array}{c}
d_0 e^{-T/\tau_a}\\
d_1 e^{-\mu \tau_a Z/2}\\
d_2\\
d_3
\end{array}
\right).
\end{equation}
Here again, the spectral parameter is identified with the pulse duration $
\lambda_a=i/\tau_a$ and  $d_0$, $d_1$, $d_2$, $d_3$ are the constants of 
integration. The simplified expressions of the elements of the involution 
matrix for large positive and negative times are shown in Table~
\ref{tab:Ma2}. 
Similarly to the other solution, the location of the initial imprint is given  
by
\begin{equation}
\kappa_a x_1= \xi_{1+}= \ln \left| \frac{d_1}{d_+}\right|.
\end{equation}

In this case, when all the coefficients are non-zero, the solution describes 
the storage of the signal pulse in the two channels. For the pulse evolution 
depicted in Fig.~(4) we only considered the storage in one channel, for which 
we took $d_3=0$. This was then superimposed with a solution consisting of 
just control pulses, in this case $d_1=0$.

For the first-order solution, the complete form for the 
fields is given by
\begin{subequations}
\label{app:comch}
\begin{align}
\Omega_1&=\frac{4 i d_0 d_1^*}{\tau_a} e^{-\kappa_a Z} g(Z,T),\\
\Omega_2&=\frac{4 i d_0 d_2^*}{\tau_a} g(Z,T),\\
\Omega_3&=\frac{4 i d_0 d_3^*}{\tau_a} g(Z,T),
\end{align}
with
\begin{equation}
g(Z,T)=\frac{e^{-T/\tau_a}}{|d_0|^2 e^{-2T/\tau_a}+|d_1|^2
e^{-2\kappa_a Z}+|d_+|^2}.
\end{equation}
\end{subequations}
It is clear that the integration with respect $T$ for the pulse areas 
only involves the function $g$ and
\begin{equation}
\int^{\infty}_{-\infty}g(Z,T) dT= \frac{\tau_a \pi}{2|d_0|}\left( 
|d_1|^2e^{-2 \kappa_a Z}+|d_+|^2 \right)^{-1/2}.
\end{equation}
The pulse area can be readily written as
\begin{subequations}
\label{app:areach}
\begin{align}
\theta_1&=\frac{2\pi i d_0 d_1^*}{|d_0|}\frac{ e^{-\kappa_a Z } }
{\sqrt{ |d_1|^2e^{-2 \kappa_a Z}+|d_+|^2 }},\\
\theta_2&=\frac{2\pi i d_0 d_2^*}{|d_0|}\frac{ 1 }
{\sqrt{ |d_1|^2e^{-2 \kappa_a Z}+|d_+|^2 }},\\
\theta_3&=\frac{2\pi i d_0 d_3^*}{|d_0|}\frac{ 1 }
{\sqrt{ |d_1|^2e^{-2 \kappa_a Z}+|d_+|^2 }}.
\end{align}
\end{subequations}
From this, the result given in Eq.~\eqref{eq:cimp1} and the three 
pulse area being equal to $2\pi$ follow.


\begin{thebibliography}{62}%
\makeatletter
\providecommand \@ifxundefined [1]{%
 \@ifx{#1\undefined}
}%
\providecommand \@ifnum [1]{%
 \ifnum #1\expandafter \@firstoftwo
 \else \expandafter \@secondoftwo
 \fi
}%
\providecommand \@ifx [1]{%
 \ifx #1\expandafter \@firstoftwo
 \else \expandafter \@secondoftwo
 \fi
}%
\providecommand \natexlab [1]{#1}%
\providecommand \enquote  [1]{``#1''}%
\providecommand \bibnamefont  [1]{#1}%
\providecommand \bibfnamefont [1]{#1}%
\providecommand \citenamefont [1]{#1}%
\providecommand \href@noop [0]{\@secondoftwo}%
\providecommand \href [0]{\begingroup \@sanitize@url \@href}%
\providecommand \@href[1]{\@@startlink{#1}\@@href}%
\providecommand \@@href[1]{\endgroup#1\@@endlink}%
\providecommand \@sanitize@url [0]{\catcode `\\12\catcode `\$12\catcode
  `\&12\catcode `\#12\catcode `\^12\catcode `\_12\catcode `\%12\relax}%
\providecommand \@@startlink[1]{}%
\providecommand \@@endlink[0]{}%
\providecommand \url  [0]{\begingroup\@sanitize@url \@url }%
\providecommand \@url [1]{\endgroup\@href {#1}{\urlprefix }}%
\providecommand \urlprefix  [0]{URL }%
\providecommand \Eprint [0]{\href }%
\providecommand \doibase [0]{http://dx.doi.org/}%
\providecommand \selectlanguage [0]{\@gobble}%
\providecommand \bibinfo  [0]{\@secondoftwo}%
\providecommand \bibfield  [0]{\@secondoftwo}%
\providecommand \translation [1]{[#1]}%
\providecommand \BibitemOpen [0]{}%
\providecommand \bibitemStop [0]{}%
\providecommand \bibitemNoStop [0]{.\EOS\space}%
\providecommand \EOS [0]{\spacefactor3000\relax}%
\providecommand \BibitemShut  [1]{\csname bibitem#1\endcsname}%
\let\auto@bib@innerbib\@empty
\bibitem [{\citenamefont {Harris}(1997)}]{harris1997electromagnetically}%
  \BibitemOpen
  \bibfield  {author} {\bibinfo {author} {\bibfnamefont {S.~E.}\ \bibnamefont
  {Harris}},\ }\href@noop {} {\bibfield  {journal} {\bibinfo  {journal} {Phys.
  Today}\ }\textbf {\bibinfo {volume} {50}},\ \bibinfo {pages} {36} (\bibinfo
  {year} {1997})}\BibitemShut {NoStop}%
\bibitem [{\citenamefont {Boller}\ \emph {et~al.}(1991)\citenamefont {Boller},
  \citenamefont {Imamo\ifmmode~\breve{g}\else \u{g}\fi{}lu},\ and\
  \citenamefont {Harris}}]{boller1991observation}%
  \BibitemOpen
  \bibfield  {author} {\bibinfo {author} {\bibfnamefont {K.-J.}\ \bibnamefont
  {Boller}}, \bibinfo {author} {\bibfnamefont {A.}~\bibnamefont
  {Imamo\ifmmode~\breve{g}\else \u{g}\fi{}lu}}, \ and\ \bibinfo {author}
  {\bibfnamefont {S.~E.}\ \bibnamefont {Harris}},\ }\href {\doibase
  10.1103/PhysRevLett.66.2593} {\bibfield  {journal} {\bibinfo  {journal}
  {Phys. Rev. Lett.}\ }\textbf {\bibinfo {volume} {66}},\ \bibinfo {pages}
  {2593} (\bibinfo {year} {1991})}\BibitemShut {NoStop}%
\bibitem [{\citenamefont {Fleischhauer}\ and\ \citenamefont
  {Lukin}(2000)}]{fleischhauer2000dark}%
  \BibitemOpen
  \bibfield  {author} {\bibinfo {author} {\bibfnamefont {M.}~\bibnamefont
  {Fleischhauer}}\ and\ \bibinfo {author} {\bibfnamefont {M.~D.}\ \bibnamefont
  {Lukin}},\ }\href@noop {} {\bibfield  {journal} {\bibinfo  {journal} {Phys.
  Rev. Lett.}\ }\textbf {\bibinfo {volume} {84}},\ \bibinfo {pages} {5094}
  (\bibinfo {year} {2000})}\BibitemShut {NoStop}%
\bibitem [{\citenamefont {Nunn}\ \emph {et~al.}(2007)\citenamefont {Nunn},
  \citenamefont {Walmsley}, \citenamefont {Raymer}, \citenamefont {Surmacz},
  \citenamefont {Waldermann}, \citenamefont {Wang},\ and\ \citenamefont
  {Jaksch}}]{nunn2007mapping}%
  \BibitemOpen
  \bibfield  {author} {\bibinfo {author} {\bibfnamefont {J.}~\bibnamefont
  {Nunn}}, \bibinfo {author} {\bibfnamefont {I.~A.}\ \bibnamefont {Walmsley}},
  \bibinfo {author} {\bibfnamefont {M.~G.}\ \bibnamefont {Raymer}}, \bibinfo
  {author} {\bibfnamefont {K.}~\bibnamefont {Surmacz}}, \bibinfo {author}
  {\bibfnamefont {F.~C.}\ \bibnamefont {Waldermann}}, \bibinfo {author}
  {\bibfnamefont {Z.}~\bibnamefont {Wang}}, \ and\ \bibinfo {author}
  {\bibfnamefont {D.}~\bibnamefont {Jaksch}},\ }\href {\doibase
  10.1103/PhysRevA.75.011401} {\bibfield  {journal} {\bibinfo  {journal} {Phys.
  Rev. A}\ }\textbf {\bibinfo {volume} {75}},\ \bibinfo {pages} {011401}
  (\bibinfo {year} {2007})}\BibitemShut {NoStop}%
\bibitem [{\citenamefont {Reim}\ \emph {et~al.}(2011)\citenamefont {Reim},
  \citenamefont {Michelberger}, \citenamefont {Lee}, \citenamefont {Nunn},
  \citenamefont {Langford},\ and\ \citenamefont {Walmsley}}]{reim2011single}%
  \BibitemOpen
  \bibfield  {author} {\bibinfo {author} {\bibfnamefont {K.~F.}\ \bibnamefont
  {Reim}}, \bibinfo {author} {\bibfnamefont {P.}~\bibnamefont {Michelberger}},
  \bibinfo {author} {\bibfnamefont {K.~C.}\ \bibnamefont {Lee}}, \bibinfo
  {author} {\bibfnamefont {J.}~\bibnamefont {Nunn}}, \bibinfo {author}
  {\bibfnamefont {N.~K.}\ \bibnamefont {Langford}}, \ and\ \bibinfo {author}
  {\bibfnamefont {I.~A.}\ \bibnamefont {Walmsley}},\ }\href {\doibase
  10.1103/PhysRevLett.107.053603} {\bibfield  {journal} {\bibinfo  {journal}
  {Phys. Rev. Lett.}\ }\textbf {\bibinfo {volume} {107}},\ \bibinfo {pages}
  {053603} (\bibinfo {year} {2011})}\BibitemShut {NoStop}%
\bibitem [{\citenamefont {Moiseev}\ and\ \citenamefont
  {Kr\"oll}(2001)}]{moiseev2001complete}%
  \BibitemOpen
  \bibfield  {author} {\bibinfo {author} {\bibfnamefont {S.~A.}\ \bibnamefont
  {Moiseev}}\ and\ \bibinfo {author} {\bibfnamefont {S.}~\bibnamefont
  {Kr\"oll}},\ }\href {\doibase 10.1103/PhysRevLett.87.173601} {\bibfield
  {journal} {\bibinfo  {journal} {Phys. Rev. Lett.}\ }\textbf {\bibinfo
  {volume} {87}},\ \bibinfo {pages} {173601} (\bibinfo {year}
  {2001})}\BibitemShut {NoStop}%
\bibitem [{\citenamefont {Kurnit}\ \emph {et~al.}(1964)\citenamefont {Kurnit},
  \citenamefont {Abella},\ and\ \citenamefont
  {Hartmann}}]{kurnit1964observation}%
  \BibitemOpen
  \bibfield  {author} {\bibinfo {author} {\bibfnamefont {N.}~\bibnamefont
  {Kurnit}}, \bibinfo {author} {\bibfnamefont {I.}~\bibnamefont {Abella}}, \
  and\ \bibinfo {author} {\bibfnamefont {S.}~\bibnamefont {Hartmann}},\
  }\href@noop {} {\bibfield  {journal} {\bibinfo  {journal} {Phys. Rev. Lett.}\
  }\textbf {\bibinfo {volume} {13}},\ \bibinfo {pages} {567} (\bibinfo {year}
  {1964})}\BibitemShut {NoStop}%
\bibitem [{\citenamefont {Mossberg}(1982)}]{mossberg1982time}%
  \BibitemOpen
  \bibfield  {author} {\bibinfo {author} {\bibfnamefont {T.~W.}\ \bibnamefont
  {Mossberg}},\ }\href@noop {} {\bibfield  {journal} {\bibinfo  {journal} {Opt.
  Lett.}\ }\textbf {\bibinfo {volume} {7}},\ \bibinfo {pages} {77} (\bibinfo
  {year} {1982})}\BibitemShut {NoStop}%
\bibitem [{\citenamefont {Gorshkov}\ \emph
  {et~al.}(2007{\natexlab{a}})\citenamefont {Gorshkov}, \citenamefont
  {Andr\'e}, \citenamefont {Fleischhauer}, \citenamefont {S\o{}rensen},\ and\
  \citenamefont {Lukin}}]{gorshkov2007universal}%
  \BibitemOpen
  \bibfield  {author} {\bibinfo {author} {\bibfnamefont {A.~V.}\ \bibnamefont
  {Gorshkov}}, \bibinfo {author} {\bibfnamefont {A.}~\bibnamefont {Andr\'e}},
  \bibinfo {author} {\bibfnamefont {M.}~\bibnamefont {Fleischhauer}}, \bibinfo
  {author} {\bibfnamefont {A.~S.}\ \bibnamefont {S\o{}rensen}}, \ and\ \bibinfo
  {author} {\bibfnamefont {M.~D.}\ \bibnamefont {Lukin}},\ }\href@noop {}
  {\bibfield  {journal} {\bibinfo  {journal} {Phys. Rev. Lett.}\ }\textbf
  {\bibinfo {volume} {98}},\ \bibinfo {pages} {123601} (\bibinfo {year}
  {2007}{\natexlab{a}})}\BibitemShut {NoStop}%
\bibitem [{\citenamefont {Gorshkov}\ \emph
  {et~al.}(2007{\natexlab{b}})\citenamefont {Gorshkov}, \citenamefont
  {Andr\'e}, \citenamefont {Lukin},\ and\ \citenamefont
  {S\o{}rensen}}]{gorshkov2007photon}%
  \BibitemOpen
  \bibfield  {author} {\bibinfo {author} {\bibfnamefont {A.~V.}\ \bibnamefont
  {Gorshkov}}, \bibinfo {author} {\bibfnamefont {A.}~\bibnamefont {Andr\'e}},
  \bibinfo {author} {\bibfnamefont {M.~D.}\ \bibnamefont {Lukin}}, \ and\
  \bibinfo {author} {\bibfnamefont {A.~S.}\ \bibnamefont {S\o{}rensen}},\
  }\href@noop {} {\bibfield  {journal} {\bibinfo  {journal} {Phys. Rev. A}\
  }\textbf {\bibinfo {volume} {76}},\ \bibinfo {pages} {033804} (\bibinfo
  {year} {2007}{\natexlab{b}})}\BibitemShut {NoStop}%
\bibitem [{\citenamefont {Gorshkov}\ \emph
  {et~al.}(2007{\natexlab{c}})\citenamefont {Gorshkov}, \citenamefont
  {Andr\'e}, \citenamefont {Lukin},\ and\ \citenamefont
  {S\o{}rensen}}]{gorshkov2007photon2}%
  \BibitemOpen
  \bibfield  {author} {\bibinfo {author} {\bibfnamefont {A.~V.}\ \bibnamefont
  {Gorshkov}}, \bibinfo {author} {\bibfnamefont {A.}~\bibnamefont {Andr\'e}},
  \bibinfo {author} {\bibfnamefont {M.~D.}\ \bibnamefont {Lukin}}, \ and\
  \bibinfo {author} {\bibfnamefont {A.~S.}\ \bibnamefont {S\o{}rensen}},\
  }\href@noop {} {\bibfield  {journal} {\bibinfo  {journal} {Phys. Rev. A}\
  }\textbf {\bibinfo {volume} {76}},\ \bibinfo {pages} {033805} (\bibinfo
  {year} {2007}{\natexlab{c}})}\BibitemShut {NoStop}%
\bibitem [{\citenamefont {Gorshkov}\ \emph
  {et~al.}(2007{\natexlab{d}})\citenamefont {Gorshkov}, \citenamefont
  {Andr\'e}, \citenamefont {Lukin},\ and\ \citenamefont
  {S\o{}rensen}}]{gorshkov2007photon3}%
  \BibitemOpen
  \bibfield  {author} {\bibinfo {author} {\bibfnamefont {A.~V.}\ \bibnamefont
  {Gorshkov}}, \bibinfo {author} {\bibfnamefont {A.}~\bibnamefont {Andr\'e}},
  \bibinfo {author} {\bibfnamefont {M.~D.}\ \bibnamefont {Lukin}}, \ and\
  \bibinfo {author} {\bibfnamefont {A.~S.}\ \bibnamefont {S\o{}rensen}},\
  }\href@noop {} {\bibfield  {journal} {\bibinfo  {journal} {Phys. Rev. A}\
  }\textbf {\bibinfo {volume} {76}},\ \bibinfo {pages} {033806} (\bibinfo
  {year} {2007}{\natexlab{d}})}\BibitemShut {NoStop}%
\bibitem [{\citenamefont {Gorshkov}\ \emph {et~al.}(2008)\citenamefont
  {Gorshkov}, \citenamefont {Calarco}, \citenamefont {Lukin},\ and\
  \citenamefont {S\o{}rensen}}]{gorshkov2008photon}%
  \BibitemOpen
  \bibfield  {author} {\bibinfo {author} {\bibfnamefont {A.~V.}\ \bibnamefont
  {Gorshkov}}, \bibinfo {author} {\bibfnamefont {T.}~\bibnamefont {Calarco}},
  \bibinfo {author} {\bibfnamefont {M.~D.}\ \bibnamefont {Lukin}}, \ and\
  \bibinfo {author} {\bibfnamefont {A.~S.}\ \bibnamefont {S\o{}rensen}},\
  }\href@noop {} {\bibfield  {journal} {\bibinfo  {journal} {Phys. Rev. A}\
  }\textbf {\bibinfo {volume} {77}},\ \bibinfo {pages} {043806} (\bibinfo
  {year} {2008})}\BibitemShut {NoStop}%
\bibitem [{\citenamefont {Lukin}\ \emph {et~al.}(2000)\citenamefont {Lukin},
  \citenamefont {Hemmer},\ and\ \citenamefont {Scully}}]{lukin2000resonant}%
  \BibitemOpen
  \bibfield  {author} {\bibinfo {author} {\bibfnamefont {M.}~\bibnamefont
  {Lukin}}, \bibinfo {author} {\bibfnamefont {P.}~\bibnamefont {Hemmer}}, \
  and\ \bibinfo {author} {\bibfnamefont {M.}~\bibnamefont {Scully}},\
  }\href@noop {} {\bibfield  {journal} {\bibinfo  {journal} {Adv. At. Mol. Opt.
  Phys.}\ }\textbf {\bibinfo {volume} {42}},\ \bibinfo {pages} {347} (\bibinfo
  {year} {2000})}\BibitemShut {NoStop}%
\bibitem [{\citenamefont {Raczy\ifmmode~\acute{n}\else \'{n}\fi{}ski}\ \emph
  {et~al.}(2004)\citenamefont {Raczy\ifmmode~\acute{n}\else \'{n}\fi{}ski},
  \citenamefont {Zaremba},\ and\ \citenamefont
  {Zieli\'{n}ska-Kaniasty}}]{raczynski2004electromagnetically}%
  \BibitemOpen
  \bibfield  {author} {\bibinfo {author} {\bibfnamefont {A.}~\bibnamefont
  {Raczy\ifmmode~\acute{n}\else \'{n}\fi{}ski}}, \bibinfo {author}
  {\bibfnamefont {J.}~\bibnamefont {Zaremba}}, \ and\ \bibinfo {author}
  {\bibfnamefont {S.}~\bibnamefont {Zieli\'{n}ska-Kaniasty}},\ }\href@noop {}
  {\bibfield  {journal} {\bibinfo  {journal} {Phys. Rev. A}\ }\textbf {\bibinfo
  {volume} {69}},\ \bibinfo {pages} {043801} (\bibinfo {year}
  {2004})}\BibitemShut {NoStop}%
\bibitem [{\citenamefont {Harris}\ and\ \citenamefont
  {Yamamoto}(1998)}]{harris1998photon}%
  \BibitemOpen
  \bibfield  {author} {\bibinfo {author} {\bibfnamefont {S.~E.}\ \bibnamefont
  {Harris}}\ and\ \bibinfo {author} {\bibfnamefont {Y.}~\bibnamefont
  {Yamamoto}},\ }\href@noop {} {\bibfield  {journal} {\bibinfo  {journal}
  {Phys. Rev. Lett.}\ }\textbf {\bibinfo {volume} {81}},\ \bibinfo {pages}
  {3611} (\bibinfo {year} {1998})}\BibitemShut {NoStop}%
\bibitem [{\citenamefont {Yan}\ \emph {et~al.}(2001)\citenamefont {Yan},
  \citenamefont {Rickey},\ and\ \citenamefont {Zhu}}]{yan2001nonlinear}%
  \BibitemOpen
  \bibfield  {author} {\bibinfo {author} {\bibfnamefont {M.}~\bibnamefont
  {Yan}}, \bibinfo {author} {\bibfnamefont {E.~G.}\ \bibnamefont {Rickey}}, \
  and\ \bibinfo {author} {\bibfnamefont {Y.}~\bibnamefont {Zhu}},\ }\href@noop
  {} {\bibfield  {journal} {\bibinfo  {journal} {Opt. Lett.}\ }\textbf
  {\bibinfo {volume} {26}},\ \bibinfo {pages} {548} (\bibinfo {year}
  {2001})}\BibitemShut {NoStop}%
\bibitem [{\citenamefont {Jiang}\ \emph {et~al.}(2006)\citenamefont {Jiang},
  \citenamefont {Chen}, \citenamefont {Zhang},\ and\ \citenamefont
  {Guo}}]{jiang2006optical}%
  \BibitemOpen
  \bibfield  {author} {\bibinfo {author} {\bibfnamefont {W.}~\bibnamefont
  {Jiang}}, \bibinfo {author} {\bibfnamefont {Q.-F.}\ \bibnamefont {Chen}},
  \bibinfo {author} {\bibfnamefont {Y.-S.}\ \bibnamefont {Zhang}}, \ and\
  \bibinfo {author} {\bibfnamefont {G.-C.}\ \bibnamefont {Guo}},\ }\href@noop
  {} {\bibfield  {journal} {\bibinfo  {journal} {Phys. Rev. A}\ }\textbf
  {\bibinfo {volume} {73}},\ \bibinfo {pages} {053804} (\bibinfo {year}
  {2006})}\BibitemShut {NoStop}%
\bibitem [{\citenamefont {Niu}\ \emph {et~al.}(2005)\citenamefont {Niu},
  \citenamefont {Gong}, \citenamefont {Li}, \citenamefont {Xu},\ and\
  \citenamefont {Liang}}]{niu2005giant}%
  \BibitemOpen
  \bibfield  {author} {\bibinfo {author} {\bibfnamefont {Y.}~\bibnamefont
  {Niu}}, \bibinfo {author} {\bibfnamefont {S.}~\bibnamefont {Gong}}, \bibinfo
  {author} {\bibfnamefont {R.}~\bibnamefont {Li}}, \bibinfo {author}
  {\bibfnamefont {Z.}~\bibnamefont {Xu}}, \ and\ \bibinfo {author}
  {\bibfnamefont {X.}~\bibnamefont {Liang}},\ }\href@noop {} {\bibfield
  {journal} {\bibinfo  {journal} {Opt. Lett.}\ }\textbf {\bibinfo {volume}
  {30}},\ \bibinfo {pages} {3371} (\bibinfo {year} {2005})}\BibitemShut
  {NoStop}%
\bibitem [{\citenamefont {Paspalakis}\ and\ \citenamefont
  {Knight}(2002)}]{paspalakis2002transparency}%
  \BibitemOpen
  \bibfield  {author} {\bibinfo {author} {\bibfnamefont {E.}~\bibnamefont
  {Paspalakis}}\ and\ \bibinfo {author} {\bibfnamefont {P.~L.}\ \bibnamefont
  {Knight}},\ }\href@noop {} {\bibfield  {journal} {\bibinfo  {journal} {J.
  Opt. B: Quantum Semiclassical Opt.}\ }\textbf {\bibinfo {volume} {4}},\
  \bibinfo {pages} {S372} (\bibinfo {year} {2002})}\BibitemShut {NoStop}%
\bibitem [{\citenamefont {Wang}\ \emph {et~al.}(2006)\citenamefont {Wang},
  \citenamefont {Marzlin},\ and\ \citenamefont {Sanders}}]{wang2006large}%
  \BibitemOpen
  \bibfield  {author} {\bibinfo {author} {\bibfnamefont {Z.-B.}\ \bibnamefont
  {Wang}}, \bibinfo {author} {\bibfnamefont {K.-P.}\ \bibnamefont {Marzlin}}, \
  and\ \bibinfo {author} {\bibfnamefont {B.~C.}\ \bibnamefont {Sanders}},\
  }\href@noop {} {\bibfield  {journal} {\bibinfo  {journal} {Phys. Rev. Lett.}\
  }\textbf {\bibinfo {volume} {97}},\ \bibinfo {pages} {063901} (\bibinfo
  {year} {2006})}\BibitemShut {NoStop}%
\bibitem [{\citenamefont {Li}\ \emph {et~al.}(2007)\citenamefont {Li},
  \citenamefont {Yang}, \citenamefont {Cao}, \citenamefont {Xie},\ and\
  \citenamefont {Wang}}]{li2007two}%
  \BibitemOpen
  \bibfield  {author} {\bibinfo {author} {\bibfnamefont {S.}~\bibnamefont
  {Li}}, \bibinfo {author} {\bibfnamefont {X.}~\bibnamefont {Yang}}, \bibinfo
  {author} {\bibfnamefont {X.}~\bibnamefont {Cao}}, \bibinfo {author}
  {\bibfnamefont {C.}~\bibnamefont {Xie}}, \ and\ \bibinfo {author}
  {\bibfnamefont {H.}~\bibnamefont {Wang}},\ }\href@noop {} {\bibfield
  {journal} {\bibinfo  {journal} {J. Phys. B: At. Mol. Opt. Phys.}\ }\textbf
  {\bibinfo {volume} {40}},\ \bibinfo {pages} {3211} (\bibinfo {year}
  {2007})}\BibitemShut {NoStop}%
\bibitem [{\citenamefont {Karpa}\ \emph {et~al.}(2008)\citenamefont {Karpa},
  \citenamefont {Vewinger},\ and\ \citenamefont {Weitz}}]{karpa2008resonance}%
  \BibitemOpen
  \bibfield  {author} {\bibinfo {author} {\bibfnamefont {L.}~\bibnamefont
  {Karpa}}, \bibinfo {author} {\bibfnamefont {F.}~\bibnamefont {Vewinger}}, \
  and\ \bibinfo {author} {\bibfnamefont {M.}~\bibnamefont {Weitz}},\
  }\href@noop {} {\bibfield  {journal} {\bibinfo  {journal} {Phys. Rev. Lett.}\
  }\textbf {\bibinfo {volume} {101}},\ \bibinfo {pages} {170406} (\bibinfo
  {year} {2008})}\BibitemShut {NoStop}%
\bibitem [{\citenamefont {Raczyński}\ \emph {et~al.}(2006)\citenamefont
  {Raczyński}, \citenamefont {Rzepecka}, \citenamefont {Zaremba},\ and\
  \citenamefont {Zielińska-Kaniasty}}]{raczynski2006polariton}%
  \BibitemOpen
  \bibfield  {author} {\bibinfo {author} {\bibfnamefont {A.}~\bibnamefont
  {Raczyński}}, \bibinfo {author} {\bibfnamefont {M.}~\bibnamefont
  {Rzepecka}}, \bibinfo {author} {\bibfnamefont {J.}~\bibnamefont {Zaremba}}, \
  and\ \bibinfo {author} {\bibfnamefont {S.}~\bibnamefont
  {Zielińska-Kaniasty}},\ }\href@noop {} {\bibfield  {journal} {\bibinfo
  {journal} {Opt. Commun.}\ }\textbf {\bibinfo {volume} {260}},\ \bibinfo
  {pages} {73 } (\bibinfo {year} {2006})}\BibitemShut {NoStop}%
\bibitem [{\citenamefont {Hang}\ and\ \citenamefont
  {Huang}(2009)}]{hang2009weak}%
  \BibitemOpen
  \bibfield  {author} {\bibinfo {author} {\bibfnamefont {C.}~\bibnamefont
  {Hang}}\ and\ \bibinfo {author} {\bibfnamefont {G.}~\bibnamefont {Huang}},\
  }\href@noop {} {\bibfield  {journal} {\bibinfo  {journal} {J. Opt. Soc. Am.
  B}\ }\textbf {\bibinfo {volume} {26}},\ \bibinfo {pages} {413} (\bibinfo
  {year} {2009})}\BibitemShut {NoStop}%
\bibitem [{\citenamefont {Si}\ \emph {et~al.}(2009)\citenamefont {Si},
  \citenamefont {Yang},\ and\ \citenamefont {Yang}}]{liu2009ultraslow}%
  \BibitemOpen
  \bibfield  {author} {\bibinfo {author} {\bibfnamefont {L.-G.}\ \bibnamefont
  {Si}}, \bibinfo {author} {\bibfnamefont {W.-X.}\ \bibnamefont {Yang}}, \ and\
  \bibinfo {author} {\bibfnamefont {X.}~\bibnamefont {Yang}},\ }\href@noop {}
  {\bibfield  {journal} {\bibinfo  {journal} {J. Opt. Soc. Am. B}\ }\textbf
  {\bibinfo {volume} {26}},\ \bibinfo {pages} {478} (\bibinfo {year}
  {2009})}\BibitemShut {NoStop}%
\bibitem [{\citenamefont {Qi}\ \emph {et~al.}(2011)\citenamefont {Qi},
  \citenamefont {Zhou}, \citenamefont {Huang}, \citenamefont {Niu},\ and\
  \citenamefont {Gong}}]{qi2011spatial}%
  \BibitemOpen
  \bibfield  {author} {\bibinfo {author} {\bibfnamefont {Y.}~\bibnamefont
  {Qi}}, \bibinfo {author} {\bibfnamefont {F.}~\bibnamefont {Zhou}}, \bibinfo
  {author} {\bibfnamefont {T.}~\bibnamefont {Huang}}, \bibinfo {author}
  {\bibfnamefont {Y.}~\bibnamefont {Niu}}, \ and\ \bibinfo {author}
  {\bibfnamefont {S.}~\bibnamefont {Gong}},\ }\href@noop {} {\bibfield
  {journal} {\bibinfo  {journal} {Phys. Rev. A}\ }\textbf {\bibinfo {volume}
  {84}},\ \bibinfo {pages} {023814} (\bibinfo {year} {2011})}\BibitemShut
  {NoStop}%
\bibitem [{\citenamefont {Petrosyan}\ and\ \citenamefont
  {Malakyan}(2004)}]{petrosyan2004magneto}%
  \BibitemOpen
  \bibfield  {author} {\bibinfo {author} {\bibfnamefont {D.}~\bibnamefont
  {Petrosyan}}\ and\ \bibinfo {author} {\bibfnamefont {Y.~P.}\ \bibnamefont
  {Malakyan}},\ }\href@noop {} {\bibfield  {journal} {\bibinfo  {journal}
  {Phys. Rev. A}\ }\textbf {\bibinfo {volume} {70}},\ \bibinfo {pages} {023822}
  (\bibinfo {year} {2004})}\BibitemShut {NoStop}%
\bibitem [{\citenamefont {Rebi\ifmmode~\acute{c}\else \'{c}\fi{}}\ \emph
  {et~al.}(2004)\citenamefont {Rebi\ifmmode~\acute{c}\else \'{c}\fi{}},
  \citenamefont {Vitali}, \citenamefont {Ottaviani}, \citenamefont {Tombesi},
  \citenamefont {Artoni}, \citenamefont {Cataliotti},\ and\ \citenamefont
  {Corbal\'an}}]{rebic2004polarization}%
  \BibitemOpen
  \bibfield  {author} {\bibinfo {author} {\bibfnamefont {S.}~\bibnamefont
  {Rebi\ifmmode~\acute{c}\else \'{c}\fi{}}}, \bibinfo {author} {\bibfnamefont
  {D.}~\bibnamefont {Vitali}}, \bibinfo {author} {\bibfnamefont
  {C.}~\bibnamefont {Ottaviani}}, \bibinfo {author} {\bibfnamefont
  {P.}~\bibnamefont {Tombesi}}, \bibinfo {author} {\bibfnamefont
  {M.}~\bibnamefont {Artoni}}, \bibinfo {author} {\bibfnamefont
  {F.}~\bibnamefont {Cataliotti}}, \ and\ \bibinfo {author} {\bibfnamefont
  {R.}~\bibnamefont {Corbal\'an}},\ }\href {\doibase
  10.1103/PhysRevA.70.032317} {\bibfield  {journal} {\bibinfo  {journal} {Phys.
  Rev. A}\ }\textbf {\bibinfo {volume} {70}},\ \bibinfo {pages} {032317}
  (\bibinfo {year} {2004})}\BibitemShut {NoStop}%
\bibitem [{\citenamefont {Ham}\ and\ \citenamefont
  {Hemmer}(2000)}]{ham2000coherence}%
  \BibitemOpen
  \bibfield  {author} {\bibinfo {author} {\bibfnamefont {B.~S.}\ \bibnamefont
  {Ham}}\ and\ \bibinfo {author} {\bibfnamefont {P.~R.}\ \bibnamefont
  {Hemmer}},\ }\href@noop {} {\bibfield  {journal} {\bibinfo  {journal} {Phys.
  Rev. Lett.}\ }\textbf {\bibinfo {volume} {84}},\ \bibinfo {pages} {4080}
  (\bibinfo {year} {2000})}\BibitemShut {NoStop}%
\bibitem [{\citenamefont {Li}\ \emph {et~al.}(2008)\citenamefont {Li},
  \citenamefont {Yang}, \citenamefont {Cao}, \citenamefont {Zhang},
  \citenamefont {Xie},\ and\ \citenamefont {Wang}}]{li2008enhanced}%
  \BibitemOpen
  \bibfield  {author} {\bibinfo {author} {\bibfnamefont {S.}~\bibnamefont
  {Li}}, \bibinfo {author} {\bibfnamefont {X.}~\bibnamefont {Yang}}, \bibinfo
  {author} {\bibfnamefont {X.}~\bibnamefont {Cao}}, \bibinfo {author}
  {\bibfnamefont {C.}~\bibnamefont {Zhang}}, \bibinfo {author} {\bibfnamefont
  {C.}~\bibnamefont {Xie}}, \ and\ \bibinfo {author} {\bibfnamefont
  {H.}~\bibnamefont {Wang}},\ }\href@noop {} {\bibfield  {journal} {\bibinfo
  {journal} {Phys. Rev. Lett.}\ }\textbf {\bibinfo {volume} {101}},\ \bibinfo
  {pages} {073602} (\bibinfo {year} {2008})}\BibitemShut {NoStop}%
\bibitem [{\citenamefont {MacRae}\ \emph {et~al.}(2008)\citenamefont {MacRae},
  \citenamefont {Campbell},\ and\ \citenamefont {Lvovsky}}]{macrae2008matched}%
  \BibitemOpen
  \bibfield  {author} {\bibinfo {author} {\bibfnamefont {A.}~\bibnamefont
  {MacRae}}, \bibinfo {author} {\bibfnamefont {G.}~\bibnamefont {Campbell}}, \
  and\ \bibinfo {author} {\bibfnamefont {A.~I.}\ \bibnamefont {Lvovsky}},\
  }\href@noop {} {\bibfield  {journal} {\bibinfo  {journal} {Opt. Lett.}\
  }\textbf {\bibinfo {volume} {33}},\ \bibinfo {pages} {2659} (\bibinfo {year}
  {2008})}\BibitemShut {NoStop}%
\bibitem [{\citenamefont {Wang}\ \emph {et~al.}(2009)\citenamefont {Wang},
  \citenamefont {Fan}, \citenamefont {Wang}, \citenamefont {Wang},
  \citenamefont {Du}, \citenamefont {Kang}, \citenamefont {Jiang},
  \citenamefont {Wu},\ and\ \citenamefont {Gao}}]{wang2009slowing}%
  \BibitemOpen
  \bibfield  {author} {\bibinfo {author} {\bibfnamefont {H.~H.}\ \bibnamefont
  {Wang}}, \bibinfo {author} {\bibfnamefont {Y.~F.}\ \bibnamefont {Fan}},
  \bibinfo {author} {\bibfnamefont {R.}~\bibnamefont {Wang}}, \bibinfo {author}
  {\bibfnamefont {L.}~\bibnamefont {Wang}}, \bibinfo {author} {\bibfnamefont
  {D.~M.}\ \bibnamefont {Du}}, \bibinfo {author} {\bibfnamefont {Z.~H.}\
  \bibnamefont {Kang}}, \bibinfo {author} {\bibfnamefont {Y.}~\bibnamefont
  {Jiang}}, \bibinfo {author} {\bibfnamefont {J.~H.}\ \bibnamefont {Wu}}, \
  and\ \bibinfo {author} {\bibfnamefont {J.~Y.}\ \bibnamefont {Gao}},\
  }\href@noop {} {\bibfield  {journal} {\bibinfo  {journal} {Opt. Lett.}\
  }\textbf {\bibinfo {volume} {34}},\ \bibinfo {pages} {2596} (\bibinfo {year}
  {2009})}\BibitemShut {NoStop}%
\bibitem [{\citenamefont {Matsko}\ \emph {et~al.}(2001)\citenamefont {Matsko},
  \citenamefont {Rostovtsev}, \citenamefont {Kocharovskaya}, \citenamefont
  {Zibrov},\ and\ \citenamefont {Scully}}]{matsko2001nonadiabatic}%
  \BibitemOpen
  \bibfield  {author} {\bibinfo {author} {\bibfnamefont {A.~B.}\ \bibnamefont
  {Matsko}}, \bibinfo {author} {\bibfnamefont {Y.~V.}\ \bibnamefont
  {Rostovtsev}}, \bibinfo {author} {\bibfnamefont {O.}~\bibnamefont
  {Kocharovskaya}}, \bibinfo {author} {\bibfnamefont {A.~S.}\ \bibnamefont
  {Zibrov}}, \ and\ \bibinfo {author} {\bibfnamefont {M.~O.}\ \bibnamefont
  {Scully}},\ }\href {\doibase 10.1103/PhysRevA.64.043809} {\bibfield
  {journal} {\bibinfo  {journal} {Phys. Rev. A}\ }\textbf {\bibinfo {volume}
  {64}},\ \bibinfo {pages} {043809} (\bibinfo {year} {2001})}\BibitemShut
  {NoStop}%
\bibitem [{\citenamefont {Shakhmuratov}\ \emph {et~al.}(2007)\citenamefont
  {Shakhmuratov}, \citenamefont {Kalachev},\ and\ \citenamefont
  {Odeurs}}]{shakhmuratov2007instantaneous}%
  \BibitemOpen
  \bibfield  {author} {\bibinfo {author} {\bibfnamefont {R.~N.}\ \bibnamefont
  {Shakhmuratov}}, \bibinfo {author} {\bibfnamefont {A.~A.}\ \bibnamefont
  {Kalachev}}, \ and\ \bibinfo {author} {\bibfnamefont {J.}~\bibnamefont
  {Odeurs}},\ }\href@noop {} {\bibfield  {journal} {\bibinfo  {journal} {Phys.
  Rev. A}\ }\textbf {\bibinfo {volume} {76}},\ \bibinfo {pages} {031802}
  (\bibinfo {year} {2007})}\BibitemShut {NoStop}%
\bibitem [{\citenamefont {Dey}\ and\ \citenamefont
  {Agarwal}(2003)}]{dey2003storage}%
  \BibitemOpen
  \bibfield  {author} {\bibinfo {author} {\bibfnamefont {T.~N.}\ \bibnamefont
  {Dey}}\ and\ \bibinfo {author} {\bibfnamefont {G.~S.}\ \bibnamefont
  {Agarwal}},\ }\href {\doibase 10.1103/PhysRevA.67.033813} {\bibfield
  {journal} {\bibinfo  {journal} {Phys. Rev. A}\ }\textbf {\bibinfo {volume}
  {67}},\ \bibinfo {pages} {033813} (\bibinfo {year} {2003})}\BibitemShut
  {NoStop}%
\bibitem [{\citenamefont {Grobe}\ \emph {et~al.}(1994)\citenamefont {Grobe},
  \citenamefont {Hioe},\ and\ \citenamefont {Eberly}}]{grobe1994formation}%
  \BibitemOpen
  \bibfield  {author} {\bibinfo {author} {\bibfnamefont {R.}~\bibnamefont
  {Grobe}}, \bibinfo {author} {\bibfnamefont {F.~T.}\ \bibnamefont {Hioe}}, \
  and\ \bibinfo {author} {\bibfnamefont {J.~H.}\ \bibnamefont {Eberly}},\
  }\href@noop {} {\bibfield  {journal} {\bibinfo  {journal} {Phys. Rev. Lett.}\
  }\textbf {\bibinfo {volume} {73}},\ \bibinfo {pages} {3183} (\bibinfo {year}
  {1994})}\BibitemShut {NoStop}%
\bibitem [{\citenamefont {McCall}\ and\ \citenamefont
  {Hahn}(1967)}]{mccall1967self}%
  \BibitemOpen
  \bibfield  {author} {\bibinfo {author} {\bibfnamefont {S.~L.}\ \bibnamefont
  {McCall}}\ and\ \bibinfo {author} {\bibfnamefont {E.~L.}\ \bibnamefont
  {Hahn}},\ }\href@noop {} {\bibfield  {journal} {\bibinfo  {journal} {Phys.
  Rev. Lett.}\ }\textbf {\bibinfo {volume} {18}},\ \bibinfo {pages} {908}
  (\bibinfo {year} {1967})}\BibitemShut {NoStop}%
\bibitem [{\citenamefont {McCall}\ and\ \citenamefont
  {Hahn}(1969)}]{mccall1969self}%
  \BibitemOpen
  \bibfield  {author} {\bibinfo {author} {\bibfnamefont {S.~L.}\ \bibnamefont
  {McCall}}\ and\ \bibinfo {author} {\bibfnamefont {E.~L.}\ \bibnamefont
  {Hahn}},\ }\href@noop {} {\bibfield  {journal} {\bibinfo  {journal} {Phys.
  Rev.}\ }\textbf {\bibinfo {volume} {183}},\ \bibinfo {pages} {457} (\bibinfo
  {year} {1969})}\BibitemShut {NoStop}%
\bibitem [{\citenamefont {Groves}\ \emph {et~al.}(2013)\citenamefont {Groves},
  \citenamefont {Clader},\ and\ \citenamefont {Eberly}}]{groves2013jaynes}%
  \BibitemOpen
  \bibfield  {author} {\bibinfo {author} {\bibfnamefont {E.}~\bibnamefont
  {Groves}}, \bibinfo {author} {\bibfnamefont {B.~D.}\ \bibnamefont {Clader}},
  \ and\ \bibinfo {author} {\bibfnamefont {J.~H.}\ \bibnamefont {Eberly}},\
  }\href@noop {} {\bibfield  {journal} {\bibinfo  {journal} {J. Phys. B: At.
  Mol. Opt. Phys.}\ }\textbf {\bibinfo {volume} {46}},\ \bibinfo {pages}
  {224005} (\bibinfo {year} {2013})}\BibitemShut {NoStop}%
\bibitem [{\citenamefont {Guti\'errez-Cuevas}\ and\ \citenamefont
  {Eberly}(2015)}]{gutierrez2015manipulation}%
  \BibitemOpen
  \bibfield  {author} {\bibinfo {author} {\bibfnamefont {R.}~\bibnamefont
  {Guti\'errez-Cuevas}}\ and\ \bibinfo {author} {\bibfnamefont {J.~H.}\
  \bibnamefont {Eberly}},\ }\href@noop {} {\bibfield  {journal} {\bibinfo
  {journal} {Phys. Rev. A}\ }\textbf {\bibinfo {volume} {92}},\ \bibinfo
  {pages} {033804} (\bibinfo {year} {2015})}\BibitemShut {NoStop}%
\bibitem [{\citenamefont {Guti\'{e}rrez-Cuevas}\ and\ \citenamefont
  {Eberly}(2015)}]{gutierrez2015multi}%
  \BibitemOpen
  \bibfield  {author} {\bibinfo {author} {\bibfnamefont {R.}~\bibnamefont
  {Guti\'{e}rrez-Cuevas}}\ and\ \bibinfo {author} {\bibfnamefont {J.~H.}\
  \bibnamefont {Eberly}},\ }\href@noop {} {\bibfield  {journal} {\bibinfo
  {journal} {J. Opt. Soc. Am. B}\ }\textbf {\bibinfo {volume} {32}},\ \bibinfo
  {pages} {2271} (\bibinfo {year} {2015})}\BibitemShut {NoStop}%
\bibitem [{\citenamefont {Clader}\ and\ \citenamefont
  {Eberly}(2007)}]{clader2007two}%
  \BibitemOpen
  \bibfield  {author} {\bibinfo {author} {\bibfnamefont {B.~D.}\ \bibnamefont
  {Clader}}\ and\ \bibinfo {author} {\bibfnamefont {J.~H.}\ \bibnamefont
  {Eberly}},\ }\href@noop {} {\bibfield  {journal} {\bibinfo  {journal} {Phys.
  Rev. A}\ }\textbf {\bibinfo {volume} {76}},\ \bibinfo {pages} {053812}
  (\bibinfo {year} {2007})}\BibitemShut {NoStop}%
\bibitem [{\citenamefont {Groves}\ \emph {et~al.}(2009)\citenamefont {Groves},
  \citenamefont {Clader},\ and\ \citenamefont {Eberly}}]{groves2009multipulse}%
  \BibitemOpen
  \bibfield  {author} {\bibinfo {author} {\bibfnamefont {E.}~\bibnamefont
  {Groves}}, \bibinfo {author} {\bibfnamefont {B.~D.}\ \bibnamefont {Clader}},
  \ and\ \bibinfo {author} {\bibfnamefont {J.~H.}\ \bibnamefont {Eberly}},\
  }\href@noop {} {\bibfield  {journal} {\bibinfo  {journal} {Opt. Lett.}\
  }\textbf {\bibinfo {volume} {34}},\ \bibinfo {pages} {2539} (\bibinfo {year}
  {2009})}\BibitemShut {NoStop}%
\bibitem [{\citenamefont {Guti\'errez-Cuevas}(shed)}]{gutierrez2016storage}%
  \BibitemOpen
  \bibfield  {author} {\bibinfo {author} {\bibfnamefont {R.}~\bibnamefont
  {Guti\'errez-Cuevas}},\ }\href@noop {} {\  (\bibinfo {year}
  {unpublished})}\BibitemShut {NoStop}%
\bibitem [{\citenamefont {Gardner}\ \emph {et~al.}(1967)\citenamefont
  {Gardner}, \citenamefont {Greene}, \citenamefont {Kruskal},\ and\
  \citenamefont {Miura}}]{gardner1967method}%
  \BibitemOpen
  \bibfield  {author} {\bibinfo {author} {\bibfnamefont {C.~S.}\ \bibnamefont
  {Gardner}}, \bibinfo {author} {\bibfnamefont {J.~M.}\ \bibnamefont {Greene}},
  \bibinfo {author} {\bibfnamefont {M.~D.}\ \bibnamefont {Kruskal}}, \ and\
  \bibinfo {author} {\bibfnamefont {R.~M.}\ \bibnamefont {Miura}},\ }\href@noop
  {} {\bibfield  {journal} {\bibinfo  {journal} {Phys. Rev. Lett.}\ }\textbf
  {\bibinfo {volume} {19}},\ \bibinfo {pages} {1095} (\bibinfo {year}
  {1967})}\BibitemShut {NoStop}%
\bibitem [{\citenamefont {Ablowitz}\ \emph {et~al.}(1973)\citenamefont
  {Ablowitz}, \citenamefont {Kaup}, \citenamefont {Newell},\ and\ \citenamefont
  {Segur}}]{ablowitz1973nonlinear}%
  \BibitemOpen
  \bibfield  {author} {\bibinfo {author} {\bibfnamefont {M.~J.}\ \bibnamefont
  {Ablowitz}}, \bibinfo {author} {\bibfnamefont {D.~J.}\ \bibnamefont {Kaup}},
  \bibinfo {author} {\bibfnamefont {A.~C.}\ \bibnamefont {Newell}}, \ and\
  \bibinfo {author} {\bibfnamefont {H.}~\bibnamefont {Segur}},\ }\href@noop {}
  {\bibfield  {journal} {\bibinfo  {journal} {Phys. Rev. Lett.}\ }\textbf
  {\bibinfo {volume} {31}},\ \bibinfo {pages} {125} (\bibinfo {year}
  {1973})}\BibitemShut {NoStop}%
\bibitem [{\citenamefont {Lamb}(1980)}]{lamb1980elements}%
  \BibitemOpen
  \bibfield  {author} {\bibinfo {author} {\bibfnamefont {G.~L.}\ \bibnamefont
  {Lamb}},\ }\href@noop {} {\emph {\bibinfo {title} {Elements of Soliton
  Theory}}}\ (\bibinfo  {publisher} {Wiley},\ \bibinfo {address} {New York},\
  \bibinfo {year} {1980})\BibitemShut {NoStop}%
\bibitem [{\citenamefont {Chakravarty}\ \emph {et~al.}(2014)\citenamefont
  {Chakravarty}, \citenamefont {Prinari},\ and\ \citenamefont
  {Ablowitz}}]{chakravarty2014inverse}%
  \BibitemOpen
  \bibfield  {author} {\bibinfo {author} {\bibfnamefont {S.}~\bibnamefont
  {Chakravarty}}, \bibinfo {author} {\bibfnamefont {B.}~\bibnamefont
  {Prinari}}, \ and\ \bibinfo {author} {\bibfnamefont {M.}~\bibnamefont
  {Ablowitz}},\ }\href@noop {} {\bibfield  {journal} {\bibinfo  {journal}
  {Physica D}\ }\textbf {\bibinfo {volume} {278}},\ \bibinfo {pages} {58}
  (\bibinfo {year} {2014})}\BibitemShut {NoStop}%
\bibitem [{\citenamefont {Lamb}(1971)}]{lamb1971analytical}%
  \BibitemOpen
  \bibfield  {author} {\bibinfo {author} {\bibfnamefont {G.~L.}\ \bibnamefont
  {Lamb}},\ }\href@noop {} {\bibfield  {journal} {\bibinfo  {journal} {Rev.
  Mod. Phys.}\ }\textbf {\bibinfo {volume} {43}},\ \bibinfo {pages} {99}
  (\bibinfo {year} {1971})}\BibitemShut {NoStop}%
\bibitem [{\citenamefont {Miura}(1976)}]{miura1976backlund}%
  \BibitemOpen
  \bibfield  {author} {\bibinfo {author} {\bibfnamefont {R.~M.}\ \bibnamefont
  {Miura}},\ }\href@noop {} {\emph {\bibinfo {title} {Backlund
  Transformations}}}\ (\bibinfo  {publisher} {Springer-Verlag},\ \bibinfo
  {address} {Berlin},\ \bibinfo {year} {1976})\BibitemShut {NoStop}%
\bibitem [{\citenamefont {Park}\ and\ \citenamefont
  {Shin}(1998)}]{park1998field}%
  \BibitemOpen
  \bibfield  {author} {\bibinfo {author} {\bibfnamefont {Q.-H.}\ \bibnamefont
  {Park}}\ and\ \bibinfo {author} {\bibfnamefont {H.~J.}\ \bibnamefont
  {Shin}},\ }\href {\doibase 10.1103/PhysRevA.57.4621} {\bibfield  {journal}
  {\bibinfo  {journal} {Phys. Rev. A}\ }\textbf {\bibinfo {volume} {57}},\
  \bibinfo {pages} {4621} (\bibinfo {year} {1998})}\BibitemShut {NoStop}%
\bibitem [{\citenamefont {Gu}\ \emph {et~al.}(2005)\citenamefont {Gu},
  \citenamefont {Hu},\ and\ \citenamefont {Zhou}}]{gu2006darboux}%
  \BibitemOpen
  \bibfield  {author} {\bibinfo {author} {\bibfnamefont {C.}~\bibnamefont
  {Gu}}, \bibinfo {author} {\bibfnamefont {H.}~\bibnamefont {Hu}}, \ and\
  \bibinfo {author} {\bibfnamefont {Z.}~\bibnamefont {Zhou}},\ }\href@noop {}
  {\emph {\bibinfo {title} {Darboux Transformations in Integrable Systems}}}\
  (\bibinfo  {publisher} {Springer},\ \bibinfo {address} {Dordrecht},\ \bibinfo
  {year} {2005})\BibitemShut {NoStop}%
\bibitem [{\citenamefont {Cie{\'s}li{\'n}ski}(2009)}]{cieslinski2009algebraic}%
  \BibitemOpen
  \bibfield  {author} {\bibinfo {author} {\bibfnamefont {J.~L.}\ \bibnamefont
  {Cie{\'s}li{\'n}ski}},\ }\href@noop {} {\bibfield  {journal} {\bibinfo
  {journal} {J. Phys. A: Math. Theor.}\ }\textbf {\bibinfo {volume} {42}},\
  \bibinfo {pages} {404003} (\bibinfo {year} {2009})}\BibitemShut {NoStop}%
\bibitem [{\citenamefont {Clader}\ and\ \citenamefont
  {Eberly}(2008)}]{clader2008two}%
  \BibitemOpen
  \bibfield  {author} {\bibinfo {author} {\bibfnamefont {B.~D.}\ \bibnamefont
  {Clader}}\ and\ \bibinfo {author} {\bibfnamefont {J.~H.}\ \bibnamefont
  {Eberly}},\ }\href@noop {} {\bibfield  {journal} {\bibinfo  {journal} {Phys.
  Rev. A}\ }\textbf {\bibinfo {volume} {78}},\ \bibinfo {pages} {033803}
  (\bibinfo {year} {2008})}\BibitemShut {NoStop}%
\bibitem [{\citenamefont {Tan-no}\ \emph {et~al.}(1972)\citenamefont {Tan-no},
  \citenamefont {Yokoto},\ and\ \citenamefont {Inaba}}]{tan1972two}%
  \BibitemOpen
  \bibfield  {author} {\bibinfo {author} {\bibfnamefont {N.}~\bibnamefont
  {Tan-no}}, \bibinfo {author} {\bibfnamefont {K.-i.}\ \bibnamefont {Yokoto}},
  \ and\ \bibinfo {author} {\bibfnamefont {H.}~\bibnamefont {Inaba}},\ }\href
  {\doibase 10.1103/PhysRevLett.29.1211} {\bibfield  {journal} {\bibinfo
  {journal} {Phys. Rev. Lett.}\ }\textbf {\bibinfo {volume} {29}},\ \bibinfo
  {pages} {1211} (\bibinfo {year} {1972})}\BibitemShut {NoStop}%
\bibitem [{\citenamefont {Eberly}\ and\ \citenamefont
  {Kozlov}(2002)}]{eberly2002wave}%
  \BibitemOpen
  \bibfield  {author} {\bibinfo {author} {\bibfnamefont {J.~H.}\ \bibnamefont
  {Eberly}}\ and\ \bibinfo {author} {\bibfnamefont {V.~V.}\ \bibnamefont
  {Kozlov}},\ }\href@noop {} {\bibfield  {journal} {\bibinfo  {journal} {Phys.
  Rev. Lett.}\ }\textbf {\bibinfo {volume} {88}},\ \bibinfo {pages} {243604}
  (\bibinfo {year} {2002})}\BibitemShut {NoStop}%
\bibitem [{\citenamefont {Shchedrin}\ \emph {et~al.}(2015)\citenamefont
  {Shchedrin}, \citenamefont {O'Brien}, \citenamefont {Rostovtsev},\ and\
  \citenamefont {Scully}}]{shchedrin2015analytic}%
  \BibitemOpen
  \bibfield  {author} {\bibinfo {author} {\bibfnamefont {G.}~\bibnamefont
  {Shchedrin}}, \bibinfo {author} {\bibfnamefont {C.}~\bibnamefont {O'Brien}},
  \bibinfo {author} {\bibfnamefont {Y.}~\bibnamefont {Rostovtsev}}, \ and\
  \bibinfo {author} {\bibfnamefont {M.~O.}\ \bibnamefont {Scully}},\
  }\href@noop {} {\bibfield  {journal} {\bibinfo  {journal} {Phys. Rev. A}\
  }\textbf {\bibinfo {volume} {92}},\ \bibinfo {pages} {063815} (\bibinfo
  {year} {2015})}\BibitemShut {NoStop}%
\bibitem [{\citenamefont {Chakravarty}(2016)}]{chakravarty2015soliton}%
  \BibitemOpen
  \bibfield  {author} {\bibinfo {author} {\bibfnamefont {S.}~\bibnamefont
  {Chakravarty}},\ }\href {\doibase
  http://dx.doi.org/10.1016/j.physleta.2015.10.031} {\bibfield  {journal}
  {\bibinfo  {journal} {Phys. Lett. A}\ }\textbf {\bibinfo {volume} {380}},\
  \bibinfo {pages} {1141 } (\bibinfo {year} {2016})}\BibitemShut {NoStop}%
\bibitem [{\citenamefont {Byrne}\ \emph {et~al.}(2003)\citenamefont {Byrne},
  \citenamefont {Gabitov},\ and\ \citenamefont
  {Kova{\v{c}}i{\v{c}}}}]{byrne2003polarization}%
  \BibitemOpen
  \bibfield  {author} {\bibinfo {author} {\bibfnamefont {J.~A.}\ \bibnamefont
  {Byrne}}, \bibinfo {author} {\bibfnamefont {I.~R.}\ \bibnamefont {Gabitov}},
  \ and\ \bibinfo {author} {\bibfnamefont {G.}~\bibnamefont
  {Kova{\v{c}}i{\v{c}}}},\ }\href@noop {} {\bibfield  {journal} {\bibinfo
  {journal} {Physica D}\ }\textbf {\bibinfo {volume} {186}},\ \bibinfo {pages}
  {69} (\bibinfo {year} {2003})}\BibitemShut {NoStop}%
\bibitem [{\citenamefont {Atkins}\ \emph {et~al.}(2012)\citenamefont {Atkins},
  \citenamefont {Kramer}, \citenamefont {Kova{\v{c}}i{\v{c}}},\ and\
  \citenamefont {Gabitov}}]{atkins2012stochastic}%
  \BibitemOpen
  \bibfield  {author} {\bibinfo {author} {\bibfnamefont {E.~P.}\ \bibnamefont
  {Atkins}}, \bibinfo {author} {\bibfnamefont {P.~R.}\ \bibnamefont {Kramer}},
  \bibinfo {author} {\bibfnamefont {G.}~\bibnamefont {Kova{\v{c}}i{\v{c}}}}, \
  and\ \bibinfo {author} {\bibfnamefont {I.~R.}\ \bibnamefont {Gabitov}},\
  }\href@noop {} {\bibfield  {journal} {\bibinfo  {journal} {Phys. Rev. A}\
  }\textbf {\bibinfo {volume} {85}},\ \bibinfo {pages} {043834} (\bibinfo
  {year} {2012})}\BibitemShut {NoStop}%
\bibitem [{\citenamefont {Newhall}\ \emph {et~al.}(2013)\citenamefont
  {Newhall}, \citenamefont {Atkins}, \citenamefont {Kramer}, \citenamefont
  {Kova{\v{c}}i{\v{c}}},\ and\ \citenamefont {Gabitov}}]{newhall2013random}%
  \BibitemOpen
  \bibfield  {author} {\bibinfo {author} {\bibfnamefont {K.~A.}\ \bibnamefont
  {Newhall}}, \bibinfo {author} {\bibfnamefont {E.~P.}\ \bibnamefont {Atkins}},
  \bibinfo {author} {\bibfnamefont {P.~R.}\ \bibnamefont {Kramer}}, \bibinfo
  {author} {\bibfnamefont {G.}~\bibnamefont {Kova{\v{c}}i{\v{c}}}}, \ and\
  \bibinfo {author} {\bibfnamefont {I.~R.}\ \bibnamefont {Gabitov}},\
  }\href@noop {} {\bibfield  {journal} {\bibinfo  {journal} {Opt. Lett.}\
  }\textbf {\bibinfo {volume} {38}},\ \bibinfo {pages} {893} (\bibinfo {year}
  {2013})}\BibitemShut {NoStop}%
\end{thebibliography}

%

\end{document}